\documentclass[fleqn,usenatbib]{mnras}

\usepackage[T1]{fontenc}
\usepackage{ae,aecompl}

\usepackage{graphicx}	
\usepackage{amsmath}	
\usepackage{amssymb}	
\usepackage{gensymb}
\usepackage{breqn}
\usepackage{color}

\newcommand{\Lim}[1]{\raisebox{0.5ex}{\scalebox{0.8}{$\displaystyle \lim_{#1}\;$}}}

\newcommand{\rg}{GM/c^2}

\title[$\kappa$monty]{$\kappa$monty: a Monte Carlo Compton Scattering code including non-thermal electrons}

\author[J. Davelaar et al.]{Jordy Davelaar$^{1,2}$\thanks{E-mail: j.davelaar@columbia.edu}, Benjamin R.~Ryan$^3$, George N.~Wong$^{4,5}$, Thomas Bronzwaer$^6$,\newauthor Hector Olivares$^6$, Monika Mo\'scibrodzka$^6$, Charles F.~Gammie$^{7,8}$, and Heino Falcke$^6$\\
$^{1}$Department of Astronomy and Columbia Astrophysics Laboratory, Columbia University, 550 W 120th St, New York, NY 10027, USA\\
$^{2}$Center for Computational Astrophysics, Flatiron Institute, 162 Fifth Avenue, New York, NY 10010, USA\\
$^{3}$CCS-2, Los Alamos National Laboratory, P.O. Box 1663, Los Alamos, NM 87545, US\\
$^{4}$School of Natural Sciences, Institute for Advanced Study, 1 Einstein Drive, Princeton, NJ 08540, USA\\
$^{5}$Princeton Gravity Initiative, Princeton University, Princeton, New Jersey 08544, USA\\
$^{6}$Department of Astrophysics/IMAPP, Radboud University, P.O. Box 9010, 6500 GL, Nijmegen, The Netherlands\\
$^{7}$Illinois Center for Advanced Studies of the Universe, Department of Physics, University of Illinois, 1110 West Green St, Urbana, IL 61801, USA\\
$^{8}$Department of Astronomy, University of Illinois, 1002 West Green Street, Urbana, IL 61801, USA
}

\date{Accepted 2023 September 29. Received 2023 September 29; in original form 2023 March 24}

\pubyear{2023}

\begin{document}
\label{firstpage}
\pagerange{\pageref{firstpage}--\pageref{lastpage}}
\maketitle

\begin{abstract}
Low-luminosity active galactic nuclei are strong sources of X-ray emission produced by Compton scattering originating from the accretion flows surrounding their supermassive black holes. The shape and energy of the resulting spectrum depend on the shape of the underlying electron distribution function (DF). In this work, we present an extended version of the {\tt grmonty} code, called {\tt $\kappa$monty}. The {\tt grmonty} code previously only included a thermal Maxwell J\"utner electron distribution function. We extend the {\tt gromty} code with non-thermal electron DFs, namely the $\kappa$ and power-law DFs, implement Cartesian Kerr-Schild coordinates, accelerate the code with {\tt MPI}, and couple the code to the non-uniform AMR grid data from the GRMHD code {\tt BHAC}. For the Compton scattering process, we derive two sampling kernels for both distribution functions. Finally, we present a series of code tests to verify the accuracy of our schemes.
The implementation of non-thermal DFs opens the possibility of studying the effect of non-thermal emission on previously developed black hole accretion models.    
\end{abstract}




\section{Introduction}

Active Galactic Nuclei (AGN) are strong sources of radiation over the full range of the electromagnetic spectrum, from radio up to $\gamma$-rays. The emission is expected to originate from a relativistic plasma flow close to these galaxies' central supermassive black holes. Low-luminosity AGN (LLAGN) are well-known sources of X-ray emission. Sagittarius A* (SgrA*), the black hole in the centre of the milky way, shows X-ray variability on the time scales of hours \citep{baganoff2003,eckart2004}. Messier 87 (M87), now famous for the first picture of a black hole shadow by the event horizon telescope collaboration (EHT) \citep{eht-paperI}, is also active in X-ray emissions \citep{Wilson2001,Marshall2002,perlman2005,prieto2016}, and shows X-ray variability on timescales of days \citep{Harris2009}. One channel to generate the X-ray emission is via inverse Compton (IC) scattering. 

To compute synthetic spectra of LLAGN, a variety of Monte Carlo codes have been developed (see e.g. \citet{yao,stern,schnittman1, schnittman2, laurent, bottcher1,bottcher2, grmonty,ryan,narayan,zhang2019,Moscibrodzka2020}). A large subset of the code uses the input from general relativistic magnetohydrodynamics (GRMHD) global simulations of weakly radiating accretion flows. In these simulations, the electron energy distribution function is not explicitly computed. GRMHD codes use a fluid approximation that only contains information on the bulk properties of the plasma and no information on the distribution function. A fluid approach also does not intrinsically contain collisionless effects. However, accretion flows in LLAGN like M87* and SgrA* have a mean free path for the electrons that is much large than the actual system size, making them effectively collisionless, and deviations from thermality are, therefore, to be expected. Magnetic reconnection, dissipation of turbulent energy, shocks, and/or other plasma instabilities that influence the shape of the electron distribution function are, in general, poorly resolved, and sub-grid models for electron heating and acceleration have to be invoked. Successful attempts to resolve magnetic reconnection in global two-dimensional GRMHD simulation have been performed by \cite{Ripperda2020, Nathanail2020} and, more recently, in three-dimensional simulations by \cite{Ripperda2022}. These models further strengthen the need for non-thermal electron distribution functions.  

One of the first attempts to model non-thermal emission from SgrA* with radiatively inefficient accretion flows (RIAF) models were made by \cite{ozel}. Later works using RIAF models include \cite{yuan2003}, among others. Semi-analytical models of RIAFs, including jets, were developed by \cite{broderick2015}, who included electron acceleration via gap acceleration. Works by \citep{Quataert2004,Chan2015a,Chan2015b} showed that for Sgr A* that X-ray bremsstrahlung emission is non-negligible. More recently, dynamical models based on GRMHD simulations were published by \cite{Chan2009,ball2016, mao2016,Chael2017,davelaar2018,davelaar2019,chatterjee2020,ACO2022,Fromm2022}. A common conclusion in all these works is that non-thermal electrons enhance the amount of NIR and radio emission. However, most of these works rely on General Relativistic Ray Tracing methods, so X-ray emission generated via IC is neither included nor approximated.

The importance of including X-ray emission in the current models used within the EHT community was shown by \cite{moscibrodzka2016}. For M87, the X-ray emission is a clear discriminator between models. If the electron temperature in the accretion disk is too high, the model easily overproduces the observed X-ray flux. However, the electron distribution function was assumed to be a thermal Maxwell-J{\"u}ttner distribution.

The X-ray emission is expected to be produced by Synchrotron Self Comptonisation. This process starts with electrons in the accretion flow that gyrate around magnetic field lines and produce emission via synchrotron emission. The emitted photons are then upscattered by the hot relativistic electrons inside the flow to X-ray and $\gamma$-ray energies via Compton scattering. Since the amount of energy that is transferred from the electron to the photon ($\Delta E$) depends on the Lorentz factor of the electron ($\gamma$) as $\Delta E \propto \gamma^2 $, the resulting spectra of the upscattered photons depend on the choice of the distribution function. It is expected that adding accelerated particles will increase the total X-ray luminosity of the source since electrons with large $\gamma$ factors and photons with larger frequencies (NIR) are present. Observed X-ray flares in AGN, and other astrophysical sources, are indicators of ongoing particle acceleration to rule out potential acceleration mechanisms models that include the generation of X-ray emission based on non-thermal electrons are needed.

In this work, we present $\kappa$monty\footnote{Publicly available at: https://github.com/jordydavelaar/kmonty} a new flavour of the Monte Carlo code {\tt grmonty} originally developed by \cite{dolence}. {\tt grmonty} is a general relativistic Monte Carlo radiative transport code developed to compute spectra of accreting black holes. A more recent version called RADPOL \citep{Moscibrodzka2020} also included polarisation and was extended to include non-thermal electron distribution functions \citep{mosc2022}. We made three large adaptations to the original \cite{dolence} code. First, we coupled our code to the non-uniform adaptive mesh refinement (AMR) grid data structure of the GRMHD code {\tt BHAC} \citep[][\url{www.bhac.science}]{porth2017,Olivares2019}, in a similar manner as we described in \cite{davelaar2019}. Second, we implemented the fit formula for the emission and absorption coefficients as obtained by \citet{pandya2016} for the initial seed photons. Third, we derived and implemented semi-analytical sampling algorithms for the $\kappa$ and power-law distribution function. The methods described in this work were used to compute the X-ray SEDs of the $\kappa$-DF based models in the Event Horizon Telescope results of Sagittarius A* \citep{ehtsgrapaperV}.

In Section \ref{sec:methods}, we explain our sampling routine and describe the setup used. In Section \ref{sec:tests}, we perform a variety of code tests. We discuss and summarise our results in Section \ref{sec:concl}.

\section{ Methods}\label{sec:methods}

In this section, we present the additions we made to the original {\tt grmonty} code \citep{grmonty}. Our new code $\kappa${\tt monty} includes the $\kappa$ distribution and power-law distribution to study accelerated particle emission and is {\tt mpi} and {\tt openmp} optimised. At first, the superphotons, a packet of photons with weight $w$, where the weight is $w$ is the number of real photons represented by the superphoton, are initialised by either thermal, $\kappa$, or power-law-based emission coefficients. As they propagate through the plasma, the total intensity decreases due to absorption. Scattering events are selected based on the mean free path length. If a photon is selected for scattering, the electron has to be drawn from the relevant distribution function. In this section, we summarise the new features of $\kappa$monty, which are new sets of coordinates, new distribution functions, and the coupling to non-uniform data formats. For a complete explanation of the initialisation, integration, and scattering of the superphotons, see the paper by \cite{grmonty}. In this section, we will give a global summary of the different aspects of the code, and we will explain in detail our modifications. 

\subsection{Geodesic integration}

The trajectory of the superphotons is computed by solving the geodesic equation,
\begin{equation}
 \frac{{\rm d}^2 x^{\alpha}}{{\rm d}\lambda^2} = -\Gamma^{\alpha}_{\ \mu \nu} \frac{{\rm d}x^{\mu}}{{\rm d}\lambda} \frac{{\rm d}x^{\nu}}{{\rm d}\lambda},
\end{equation}
where $\Gamma^{\alpha}_{\ \mu\nu}$ are the Christoffel symbols, and $\lambda$ the affine parameter. The Christoffel symbols depend on derivates of the metric and are given by,
\begin{equation}
 \Gamma^{\alpha}_{\ \mu \nu} = \frac{1}{2} g^{\alpha \rho} \left[
 \partial_{\mu} g_{\nu \rho} + \partial_{\nu} g_{\mu \rho} -
 \partial_{\rho} g_{\mu \nu} \right].
 \label{eqn:christoffels}
\end{equation}
The Christoffel symbols are either provided analytically or can be computed by computing the metric derivatives using a second-order finite difference method. 

\subsubsection{Kerr-Schild coordinates}
A rotating black hole is described by the Kerr metric \citep{kerr1963}. In spherical Kerr-Schild horizon penetrating coordinates the non-zero covariant components of the metric\footnote{We use the metric signature (-,+,+,+).} $g_{\mu\nu}$ are given by
\begin{subequations}
\begin{align}
g_{\rm tt} &= - \left(1 - \frac{2r}{\Sigma}\right) \\
g_{{\rm t}\phi} &= g_{\phi{\rm t}} = - \frac{2  r a_* \sin^2 \theta}{\Sigma}\\
g_{\rm rr} &= \frac{\Sigma}{\Delta}\\
g_{\theta\theta} &= \Sigma\\
g_{\phi\phi} &= \left(r^2 + a_*^2 + \frac{2  r a_*^2}{\Sigma}\sin^2\theta \right) \sin^2 \theta
\end{align}
\end{subequations}
for a black hole with unitary mass $M=1$ and angular momentum $J$,
where $a_*={J}/(Mc)$ is the dimensionless spin parameter,  $\Sigma = r^2 + a_*^2 \cos^2 \theta$, and $\Delta = r^2 - 2  r + a_*^2$. 

The GRMHD code used in this work {\tt BHAC} primarily uses for the EHT GRMHD library \citep{eht-paperV} Modified Kerr-Schild coordinates where the $r$ and $\theta$ coordinates are modified. The coordinates $(t,X_1,X_2,X_3)$ are related to standard KS via

\begin{subequations}
\begin{align}
t&=t\\
r&= \exp(X_1) \\
\theta&= X_2 + \frac{h}{2} \sin(2 X_2)\\
\phi &= X_3
\end{align}
\end{subequations}
This results in a grid that is logarithmic spaced in radius and concentrated towards the midplane in $\theta$, set by the $h$ parameter. Transforming the metric to MKS is done via multiplication of the metric terms with the  non-zero elements of the Jacobian,
\begin{subequations}
\begin{align}
\partial r/\partial x_1 &= r \\
\partial \theta/\partial X_2 &= 1 + h \cos(2 X_2).
\end{align}
\end{subequations}

\subsubsection{Cartesian Kerr-Schild coordinates}
We extended the code to include Cartesian Kerr-Schild (CKS) coordinates, which relate to spherical Kerr-Schild coordinates via

\begin{subequations}
\begin{align}
t&=t\\
x&=r (\cos(\hat{\phi}) + a_* \sin(\hat{\phi}))\sin(\theta)\\
y&=r (\sin(\hat{\phi}) - a_* \cos(\hat{\phi}))\sin(\theta)\\
z&= r \cos(\theta).
\end{align}
\end{subequations}

The covariant Cartesian KS metric, $g_{\mu\nu}$, is given by \citep{kerr1963}
\begin{equation}
g_{\mu\nu} = \eta_{\mu\nu} + f l_\mu l_\nu,
\end{equation}
here $\eta_{\mu\nu}$ is the Minkowski metric and is given by ${\eta_{\mu\nu}={\rm diag}{(-1,1,1,1)}}$, and
\begin{subequations}
\begin{align}
f &= \frac{2r^3}{r^4 + a_*^2 z^2},\\
l_\nu &= \left(1, \frac{rx+a_* y}{r^2 + a^2}, \frac{ry-a_* x}{r^2 + a_*^2}, \frac{z}{r}\right),
\end{align}
\end{subequations}
where $r$ is given by
\begin{equation}
r^2 = {\frac{R^2 - a_*^2 + \sqrt{(R^2 - a_*^2)^2 + 4a_*^2z^2}}{2} },
\end{equation}
and 
\begin{eqnarray}
R^2 = x^2+y^2+z^2.
\end{eqnarray}
In the limit of $R \gg a_*$, the radius $r
\rightarrow R$. The contravariant metric is defined as
\begin{equation}
g^{\mu\nu} = \eta^{\mu\nu} - f l^\mu l^\nu,
\end{equation}
where $l^\nu$ is given by
\begin{eqnarray}
l^\nu = \left(-1, \frac{rx+a_*y}{r^2 + a_*^2}, \frac{ry-a_*x}{r^2 + a_*^2}, \frac{z}{r}\right).
\end{eqnarray}

\subsection{Distribution functions}

\begin{figure}
  \centering
  \includegraphics[width=0.5\textwidth]{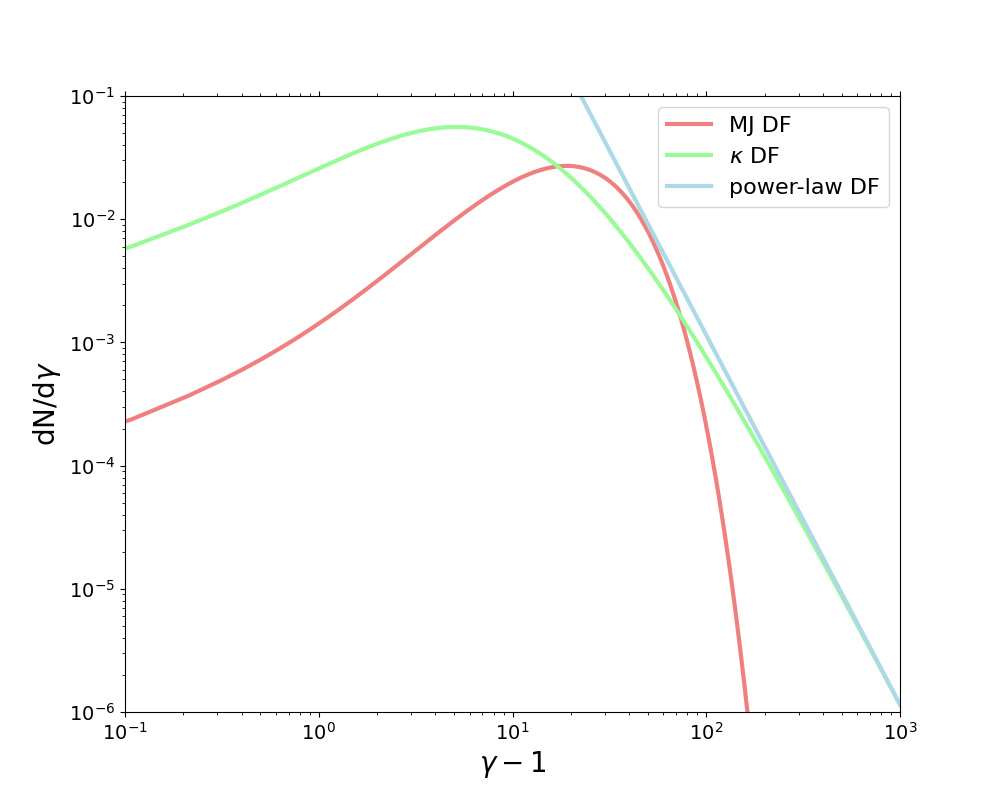}
  \caption{The Maxwell-J\"uttner (MJ), $\kappa$ and power-law distribution functions (DF). The MJ DF is shown for a dimensionless electron temperature of $\Theta_{\rm e}=10$, the $\kappa$ DF is shown with $\kappa=4.0$, and $w = 2.5$, and the power-law DF is shown with $p=3$, $\gamma_{\rm min}=25$, and $\gamma_{\rm max}=10^3$.}
    \label{fig:dfs}

  \end{figure}

For the distribution function, we either use a Maxwell-J{\"u}ttner (MJ) distribution function (DF), a $\kappa$-DF, or a power law-DF. All three distribution functions are isotropic. 

The MJ DF is given by
\begin{equation}
\frac{dn_e}{d\gamma} = \frac{n_e}{\Theta_e} \frac{\gamma^2 \beta}{K_2(\Theta_e^{-1})}\exp{\left(-\frac{\gamma}{\Theta_e}\right)}\,,
\end{equation}
where $\gamma_e$ is the Lorentz factor of the electrons, $n_e$ the number density of electrons, $\Theta_e$ the dimensionless electron temperature, and $K_2$ the modified Bessel function of the second kind. For the thermal DF, the emission coefficients used in {\tt $\kappa$monty} can be found in \cite{leung}.

The $\kappa$ DF is used to describe the particle population of a variety of space plasma, such as the solar wind \citep{decker}, coronal flares on the Sun \citep{Livadiotis2013}, turbulent flows \citep{kunz2016}, and jets \citep{davelaar2018}. X-ray spectra generated based on this distribution function could, therefore, be of interest to a broad range of astrophysical problems. The DF in relativistic form \citep{xiao2006} is given by,
\begin{equation}
\frac{dn_e}{d\gamma} = n_e N \gamma \sqrt{\gamma^2 - 1 }\left(1+\frac{\gamma-1}{\kappa w}\right)^{-(\kappa +1)},
\end{equation}
where the $\kappa$-parameter sets the power law-index via $\kappa=p+1$, $w$ is the width of the distribution function, and $N$ is a normalisation constant. In the $\kappa$ case, the normalisation constant $N$ is not known analytically and is, therefore, when needed, computed during run time with a {\tt gsl} integrator by demanding that
\begin{equation}
  \int_1^\infty \frac{dn_e}{n_e d\gamma} d\gamma = 1.
\end{equation}
The emission coefficients for the $\kappa$-DF can be found in \cite{pandya2016}.

Finally, the power-law DF is given by
\begin{equation}
    \frac{dn_e}{d\gamma} = n_e \frac{(p-1)\gamma^{-p}}{(\gamma_{\rm min}^{1-p} - \gamma_{\rm max}^{1-p})},
\end{equation}
where $p$ is the power-law index. The DF function is non-zero only when $\gamma$ is between $\gamma_{\rm min}$ and $\gamma_{\rm max}$. The emission coefficients for the power law-DF can be found in \cite{pandya2016}.

All three DFs are shown in figure \ref{fig:dfs}. The dependence of the DF in the code can be found in three places; the emission coefficients, the cross-section for scattering, and the sampling of the DF if a scattering event takes place. In the remainder of this section, we will explain what changes we made to the code for each of these.

\subsection{Emission coefficients}
For the emission coefficients, the code uses fit formulas from \cite{leung2011} for the thermal distribution function and $\kappa$ and power-law from \cite{pandya2016}. The fit formulas for the $\kappa$ coefficients are only valid for $\kappa<7.5$, they do not recover the thermal DF in the limit of $\kappa \rightarrow \infty$. 

\subsection{Cross-sections}

The cross-section for an IC scattering is dependent on the local electron population, both the energy budget as well as the shape of the distribution. The cross-section is given by
\begin{equation}
  \alpha_\nu = n_m \sigma_h
\end{equation}
where $\sigma_h$ is defined as the ``hot cross section'',
\begin{equation}
  \sigma_h = \frac{1}{n_e} \int d^3p \frac{dn_e}{d^3p}(1-\mu_e\beta_e)\sigma,
\end{equation}
where $n_e$ is the number density of electrons, $\frac{dn_e}{d^3p}$ is the electron distribution function, $\mu_e$ is the cosine of the angle between the superphoton momentum and the electron momentum, and $\beta_e$ is the electron speed in the plasma frame, and $\sigma$ the Klein-Nishina total cross-section. 

From the cross-section the code computes the scattering opacity $\tau_{\rm s}$, via $\tau_{\rm s} = \alpha_{\rm s} r_{\rm g} h / (m_{\rm e} c^2) \Delta \lambda$, where $\alpha_{\rm s}$ is the extinction coefficient given by $\alpha_{\rm s} = n_{\rm e} \sigma_h$. The total probability for a scattering event is then given by $p=1 - e^{- b \tau_{\rm s}}$, where $b$ is a bias factor that enhances the scattering probability, as introduced in \cite{grmonty}. This bias is then counteracted by splitting the scattered superphoton into an upscattered superphoton with weight $bw$ and an unscattered remnant superphoton with weight $(1 - b ) w$. In this work, $\kappa${\tt monty} only exploits the original {\tt grmonty} bias function given by $b= \Theta_{\rm e} / \langle \Theta_{\rm e} \rangle$, where $\langle \Theta_{\rm e} \rangle$ is the volume averaged dimensionless electron temperature. More fine-tuned bias functions, such as the one in {\tt igrmonty} \citep{wong2022}, are not explored since the main focus of this work is the numerical algorithms for non-thermal DF sampling.

\subsection{Sampling routines}
The outcome of a Compton scattering event between a superphoton, and an electron depends on the superphoton's wavevector and the electron's four-velocity. The electrons are, in the case of GRMHD models, coupled to the plasma parameters of the protons, which are approximated by a fluid description. We, therefore, only know ensemble averages. To be able to select a single electron, we need a sampling algorithm that, given a set of plasma variables, draws a $\gamma$ factor based on the chosen DF, in the original {\tt grmonty}, the procedure from \citet{canfield} is used for the MJ DFs. Therefore, only new samplers for the $\kappa$-DF and power law-DF are needed. 

\subsubsection{A semi-analytical sampling routine for the $\kappa$ distribution function}
For the $\kappa$-DF, we will generalise the procedure from  \citet{canfield} for the $\kappa$-distribution function. The relativistic $\kappa$-distribution function as function of velocity $\beta=\frac{v}{c}$ is given by
\begin{equation}
 f_\kappa ( \beta,w) = N w \beta^2 \gamma^5 \left(1+\frac{\gamma-1}{\kappa w}\right)^{-\kappa -1},
\end{equation}
with $\gamma = (1-\beta^2)^{-1/2}$.

To sample electrons based on this distribution function, we derive a Monte Carlo-based scheme. We first introduce a random variable $y$ that is coupled to $\gamma$ and $w$
\begin{equation}
 y^2 = \frac{\gamma - 1}{w},
\end{equation}
and transform our probability density function (pdf) accordingly
\begin{align}
 f_\kappa ( y,w)= & f_\kappa ( \beta,w) \frac{\partial \beta}{\partial y},\\
 f_\kappa ( y,w) = & Nw \sqrt{2w}y^2 \sqrt{1+0.5wy^2}(1+wy^2)\left(1+\frac{y^2}{\kappa }\right)^{-\kappa -1}.
\end{align}
Following the procedure by \citet{canfield}, we can split our pdf into a series of pdfs after multiplying with
\begin{align}
 {1} = & \frac{1+\sqrt{0.5w}y}{1+\sqrt{0.5w}y},
\end{align}
to obtain
\begin{align}
 f_\kappa ( y,w) =  \sqrt{2w}y^2 \left(1+\sqrt{0.5w}y\right)\left(1+wy^2\right) \times \notag\\ \left(1+\frac{y^2}{\kappa }\right)^{-\kappa -1} Nw \frac{\sqrt{1+0.5wy^2}}{1+\sqrt{0.5w}y}, 
\end{align}
this can be rewritten as
\begin{align}
 f_\kappa ( y,w) =  \left(y^2 + \sqrt{0.5w}y^3 + w y^4 + w\sqrt{0.5w}y^5\right) \times \notag\\ \left(1+\frac{y^2}{\kappa }\right)^{-\kappa -1} N w\sqrt{2w} \frac{\sqrt{1+0.5wy^2}}{1+\sqrt{0.5w}y}.
\end{align}

We can now identify two different functions that are the core of the sampling routine, a rejection function $H_3(w,y)$ and a sampling function $G_3(w,y)$ such that
\begin{equation}
 f_\kappa ( y,w) = G_3(w,y) H_3(w,y).
\end{equation}
The sampling function is given by
\begin{equation}
 G_3(w,y) =  \sum_{j=3}^6 \pi_j(w) g_j(y),
\end{equation}
and consists of a series of sample coefficients $g_j(y)$ and probability coefficients $\pi_j(w)$. The sampling coefficients are given by

\begin{equation}
g_j(y) \equiv \frac{y^{j-1}}{n_j} \left(1 + \frac{y^2}{\kappa} \right)^{-\kappa-1}    
\end{equation}

where $n_j$ is a normalisation constant obtained by integrating
\begin{equation}
 \int_0^\infty n_j g_j(y) dy = n_j.
\end{equation}
Performing these four integrals, we get
\begin{eqnarray}
 n_3 = \frac{\sqrt{\kappa}\sqrt{\pi}\Gamma(-\frac{1}{2} + \kappa)}{4 \Gamma(\kappa)},\\
 n_4= \frac{\kappa}{2(-1+\kappa)},\\
 n_5= \frac{3{\kappa}^{3/2}\sqrt{\pi}\Gamma(-\frac{3}{2} + \kappa)}{8 \Gamma(\kappa)},\\
 n_6= \frac{\kappa^2}{2-3\kappa+\kappa^2},
\end{eqnarray}
where $\Gamma(x)$ is the Gamma function. The analytical solutions for the normalisations are only valid in the case that $\kappa>2$. The probability coefficients are given by
\begin{eqnarray}
 \pi_3(w) = \frac{n_3}{S_3(w)} \\
 \pi_4(w) = \frac{n_4 \sqrt{0.5w}}{S_3(w)}\\
 \pi_5(w) = \frac{n_5 w}{S_3(w)}\\
 \pi_6(w) = \frac{n_6 w\sqrt{0.5w}}{S_3(w)}
\end{eqnarray}
with
\begin{equation}
 S_3(w) = n_3 + n_4 \sqrt{0.5w} + n_5 w + n_6 w\sqrt{0.5w}.
\end{equation}
The rejection function $H_3(w,y)$ is then defined as
\begin{equation}
 H_3(w,y)= \frac{N t\sqrt{2w}}{S_3(w)} \frac{\sqrt{1+0.5wy^2}}{1+\sqrt{0.5w}y}.
\end{equation}
The rejection criterion is then, similarly to \citet{canfield},
\begin{equation}
 h(w,y) =  \frac{\sqrt{1+0.5wy^2}}{1+\sqrt{0.5w}y}.
\end{equation}
The rejection criterion can be generalised even more when one also wants to add an exponential cutoff to the $\kappa$-distribution function,
\begin{equation}
 f_{\kappa,\gamma_{\rm cutoff}}(\gamma) =  f_{\kappa}(\gamma) e^{-\frac{\gamma}{\gamma_{\rm cutoff}}}.
\end{equation}
We can contract the exponential cutoff into the rejection criterion. Since if we encounter a large value of $\gamma$, it will decrease the likelihood of being accepted by the sampling routine,
\begin{equation}
 h(w,y) =  \frac{\sqrt{1+0.5wy^2}}{1+\sqrt{0.5w}y} e^{-\frac{w  y^2}{\gamma_{\rm cutoff}}}.
\end{equation}

The procedure for sampling the $\kappa$ distribution function is, therefore
\begin{enumerate}
  \item Draw a random number $x_1$
  \item If $x_1$ < $\pi_j$
  \item Find $y$ according to $g_j(y)$
  \item Draw a random number $x_2$
  \item Accepted $y$ when $x_2<h(w,y)$
  \end{enumerate}
This rejection constraint is, as mentioned in \citet{canfield}, very efficient because for large values of $w$ or small values of $w$, $h(w,y)$ asymptotes to one.\\

 The last step in this derivation of the sampling routine is to find a procedure for the third step, find  $y$ according to $g_j(y)$. In the case of a thermal distribution function, the $g_j(y)$ are $\chi^2$ functions that can be sampled with standard {\tt gsl} library functions. In the case of the $\kappa$-distribution function, this is less straightforward. To sample $g_j(y)$, we make use of the fact that the cumulative distribution functions (CDF) belonging to the pdfs $g_j(y)$ are monotonically increasing functions between zero and one. These CDFs can be obtained by integrating $g_j(y)$ from zero to $y$,
 \begin{equation}
  F_j(y) =  \int_0^y g_j(y) dy.
 \end{equation}
 Performing these four integrals result in
 \begin{align}
  F_3(y) = & -\frac{\sqrt{\kappa} \left(\frac{y^2+\kappa}{\kappa}\right)^{-\kappa}
   \Gamma (\kappa)}{\sqrt{\pi } y \Gamma \left(\kappa+\frac{3}{2}\right)} \times \notag \\  & \left[-\kappa  _2F_1\left(1,-\kappa-{1}/{2};{1}/{2};-{y^2}/{\kappa}\right)+y^2 (2
   \kappa+1)+\kappa\right], 
 \end{align}
 \begin{dmath}
  F_4(y) = 1-\left(y^2+1\right) \left(\frac{y^2+\kappa}{\kappa}\right)^{-\kappa},\\
 \end{dmath}
 \begin{align}
  F_5(y) = & \frac{\left(\frac{y^2+\kappa}{\kappa}\right)^{-\kappa}  \Gamma (\kappa)}{3 \sqrt{\pi } y \sqrt{\kappa} \Gamma \left(\kappa+\frac{3}{2}\right)} \notag\\ 
  & \left[ 3 \kappa^2 \left(\, _2F_1\left(1,-\kappa-{1}/{2};{1}/{2};-{y^2}/{\kappa}\right)-1\right) \right.  + \notag\\ 
  & \left. \left(1-4 \kappa^2\right) y^4-3 \kappa (2 \kappa+1) y^2\right],
 \end{align}
 \begin{dmath}
  F_6(y) = \frac{\left(y^4-\left(y^4+2 y^2+2\right) \kappa\right)
   \left(\frac{y^2+\kappa}{\kappa}\right)^{-\kappa}}{2 \kappa}+1,
 \end{dmath}
where $_2F_1$ is the second-order hypergeometrical function of the first kind. We can then find an $y$ by using an inverse transform sampling method,

\begin{enumerate}
  \item draw a number $u$ from a uniform distribution $[0,1]$
  \item solve such that $F_j(y)=u$, where  $F_j(y)$ is the  cumulative distribution function of $g_j(y)$
  \item $y$ is sampled according to $g_j(y)$
\end{enumerate}
To solve step two, we implemented a Brent root-finding algorithm.

For the implementation of the algorithm, special attention has to be paid to the hypergeometrical functions encountered in the CDFs. For negative integer values of the arguments of the hypergeometrical function, it is impossible to use the series expansion form, as implemented in the { \tt gsl} library. We, therefore, pre-computed a table of the hypergeometrical function as a function of $\kappa$ and $y$ with {\tt Mathematica}, which is read in by $\kappa$monty. The resulting table is then interpolated with a first-order interpolation scheme.

\subsubsection{The \texorpdfstring{$\kappa \rightarrow \infty$}{kappa to infinity} limit}
 In the case that $\kappa \rightarrow \infty$ we expect our derived sampler to recover the original sampler by \cite{canfield}.
 
 First, we check that eqn. \ref{eqn-pdfs} recovers the pdfs in \cite{canfield} by taking $\Lim{{\kappa \rightarrow \infty}} g_j(y)$, resulting in,
\begin{eqnarray}
 g_3(y)= \lim_{\kappa \rightarrow \infty} \frac{y^2}{n_3} \left(1+\frac{y^2}{\kappa }\right)^{-\kappa -1} = \frac{y^2}{n_3} e^{y^2}  \\
 g_4(y)= \lim_{\kappa \rightarrow \infty} \frac{y^3}{n_3} \left(1+\frac{y^2}{\kappa }\right)^{-\kappa -1} = \frac{y^3}{n_4} e^{y^2} \\
 g_5(y)= \lim_{\kappa \rightarrow \infty} \frac{y^4}{n_4} \left(1+\frac{y^2}{\kappa }\right)^{-\kappa -1} = \frac{y^4}{n_5} e^{y^2} \\
 g_6(y)= \lim_{\kappa \rightarrow \infty} \frac{y^5}{n_5} \left(1+\frac{y^2}{\kappa }\right)^{-\kappa -1} = \frac{y^5}{n_6} e^{y^2}.
\end{eqnarray}
 Here we used that $\Lim{{\kappa \rightarrow \infty}} (1+y/\kappa)^{-(\kappa+1)} = e^y$. The resulting $g_j(y)$ in this limit are consistent with \cite{canfield}.
 
 Secondly, we check $\Lim{{\kappa \rightarrow \infty}} n_j(y)$, which are given by 
 \begin{eqnarray}
 n_3 = \lim_{\kappa \rightarrow \infty}\frac{\sqrt{\kappa}\sqrt{\pi}\Gamma(-\frac{1}{2} + \kappa)}{4 \Gamma(\kappa)} = \sqrt{\pi}/4, \\
 n_4=\lim_{\kappa \rightarrow \infty} \frac{\kappa}{2(-1+\kappa)}=1/2,\\
 n_5=\lim_{\kappa \rightarrow \infty} \frac{3{\kappa}^{3/2}\sqrt{\pi}\Gamma(-\frac{3}{2} + \kappa)}{8 \Gamma(\kappa)}=3\sqrt{\pi}/8,\\
 n_6=\lim_{\kappa \rightarrow \infty} \frac{\kappa^2}{2-3\kappa+\kappa^2}=1.
\end{eqnarray}
 Here we use that $\Lim{{\kappa \rightarrow \infty}} \kappa^{n/2} \Gamma(-n/2+\kappa)/\Gamma(\kappa) = 1$. The resulting formulas are consistent with \cite{canfield}. Our last test is to check the rejection criterion, which is already in the same form and is equal to the one by \cite{canfield} in the case that $\Lim{\kappa \rightarrow \infty} w = \Theta_e$, which is the case if $w=\frac{\kappa-3}{\kappa} \Theta_{\rm e}$.
 
 \subsubsection{A numerical sampling routine for the $\kappa$ distribution function}
 
 We also implemented a more mundane rejection sampling method for drawing electrons from the $\kappa$ distribution. This implementation can be found in the public code {\tt igrmonty} \footnote{https://github.com/AFD-Illinois/igrmonty} as well as in {\tt $\kappa$monty}. The expression $\partial f_{\kappa} / \partial \gamma = 0$ is solved for $\gamma$, hereafter $\gamma_{\rm max}$. Fiducial $\gamma$ are then drawn uniformly in log space between 
 \begin{eqnarray}
 \gamma_{\rm min}, = {\rm MAX}(1, 0.01\times \Theta_e) \\
 \gamma_{\rm max}, = {\rm MAX}(100, 1000\times\Theta_e)
 \end{eqnarray}
 where these parameters are chosen to ensure both accuracy and computational efficiency for all $\Theta_e$. New $\gamma$ are drawn until the condition $\gamma f_{\kappa}(\gamma)/\gamma_{\rm max} f_{\kappa}(\gamma_{\rm max}) > {\rm rand}$, where ${\rm rand}$ is a uniformly distributed random number in the range $[0, 1)$ and the extra factors of $\gamma$ arise from drawing fiducial $\gamma$ uniformly in log space. 
 
 This procedure generalises to any realistic electron distribution function, including the MJ distribution, but also, e.g., anisotropic DFs, DF based on charged test particles in MHD, or DFs based on first-principle PIC simulations. Compared to the \cite{canfield} prescription for sampling MJ, this rejection sampling approach leads to only modestly ($\sim$ 20 \%) slower calculation wallclock times.

\subsubsection{Power-law distribution}

For the power-law distribution function, the procedure is much more trivial. The cumulative distribution function is, in this case, given by
\begin{equation}\label{CDF-pwl}
F(\gamma) = \int_{\gamma_{\rm min}}^\gamma \frac{(p-1)\gamma^{-p}}{(\gamma_{\rm min}^{1-p} - \gamma_{\rm max}^{1-p})} d\gamma =\frac{\left(\gamma_{\rm min}^{1-p} - \gamma^{1-p}\right)}{\left(\gamma_{\rm min}^{1-p} - \gamma_{\rm max}^{1-p}\right)}.   
\end{equation}
Since this CDF can be inverted analytically, we can use an inverse sampling method. We first pick a random number $x$ between zero and one, and $\gamma$ can then be found by computing the inverse of eqn. \ref{CDF-pwl} and setting $F(\gamma)=x$,
\begin{equation}
    \gamma = \left((1- x)\gamma_{\rm min}^{1-p} + x \gamma_{\rm max}^{1-p}\right)^{1/(1-p)}.
\end{equation}
\subsection{Interface with BHAC}

The Black Hole Accretion Code ({\tt BHAC}) \citep{porth2017,Olivares2019} is a finite volume code that solves the covariant GRMHD equation using a 3+1 split. The code is capable of using AMR grids that, during runtime, based on user-defined criteria, can refine or derefine the grid. {\tt BHAC} outputs the GRMHD data in an octree structure. We fully interfaced $\kappa${\tt monty} to this data format similarly to \cite{davelaar2019}, and made $\kappa${\tt monty} capable of initialising superphotons and performing IC in a non-uniform grid.

\section{Code verification}\label{sec:tests}
 To verify our implementation of the described methods, we performed extensive code tests. In this section, we describe the performed tests and discuss the results.
 
 \subsection{Sampling routines}
\begin{figure}
  \centering
  \includegraphics[width=0.5\textwidth]{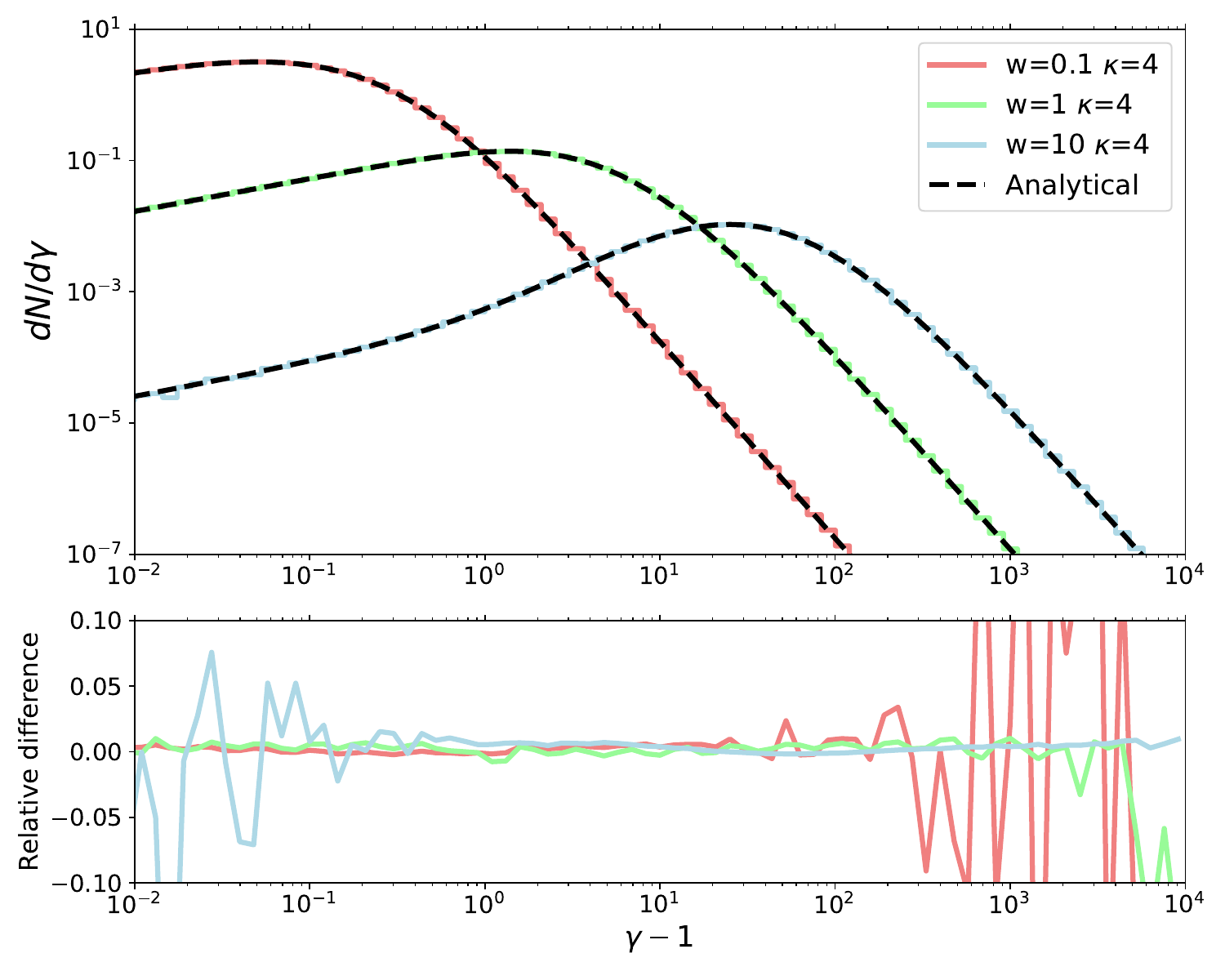}
  \caption{Output of the $\kappa$ sampler compared to the analytical form for: $w=0.1$ (red), $w=1.0$ (green) and $w=10$ (blue). All three cases use $\kappa=4.0$. The majority of the DF from the sampler shows almost perfect agreement with respect to the exact form. Only at a low or high Lorentz factor the error increases for the $w=10$ and $w=0.1$ cases, respectively. This is caused by the MC nature of our sampler, which makes regions with small electron number density harder to sample due to MC noise. }
    \label{fig:ksampler}

  \end{figure}

\begin{figure}
  \centering
   \includegraphics[width=0.5\textwidth]{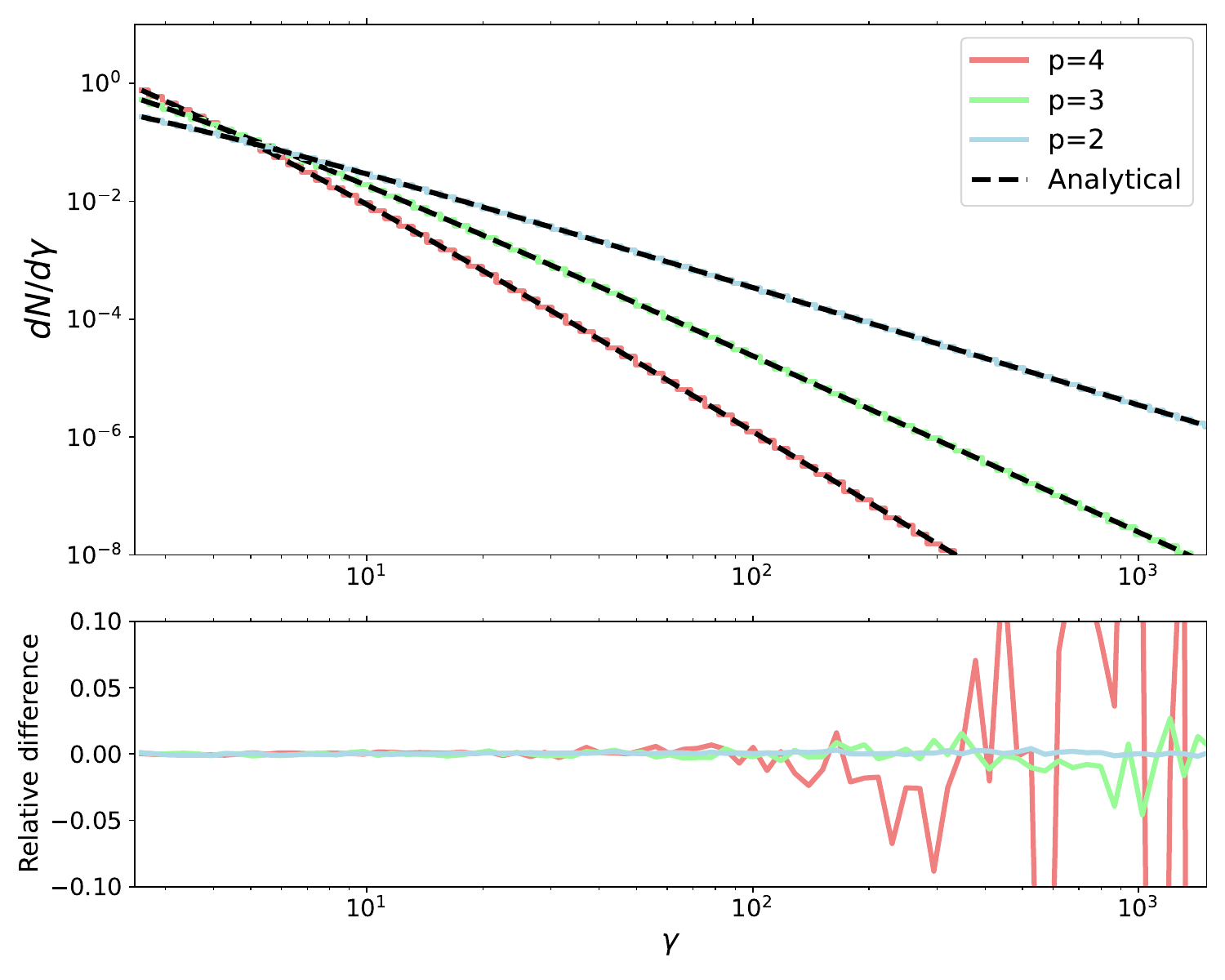}
  \caption{Output of the power-law sampler compared to the analytical form, $p=4.0$ (red), $p=3.0$ (green), and $p=2.0$ (blue). For all three cases we set $\gamma_{\rm min}=1.0$ and $\gamma_{\rm max}=10^3$. The deviation between the sampler and the exact form is for almost the entire range less than 1\%, except for large $\gamma$ values, due to the steep power-law populating the high end of the DF is affected by MC noise.}
  \label{fig:psampler}
 \end{figure}

We first tested the sampling routines by computing $\kappa$-distribution functions for three values of $w=[0.1,1,10]$ and $\kappa=4$. This is done by drawing $10^9$ electrons and comparing the resulting distribution with the analytical form. The results of this are shown in figure \ref{fig:ksampler}. Overall the relative difference between the exact form and the distribution is close to 0\% for all values of $w$, except for $\gamma$ values close to one for $w=10$, or large $\gamma$ values in the case of $w=0.1$. For these $\gamma$ values, the distribution functions have a small population of electrons, which are therefore dominated by MC noise. Similarly, for the power-law distribution, we computed distribution for three values of the power-law index, $p=[3,4,5]$ and $\gamma_{\rm min}=3.5$, and $\gamma_{\rm max}=10^4$. The results of this are shown in figure \ref{fig:psampler}. We find good agreement between the analytical form and the output of our sampler, and large deviations from 0\% are only found in regions with low electron number densities, e.g., large $\gamma$ values. 

\subsection{Uniform sphere test}

To test the implementation of both the emission coefficients and the scattering kernel, we designed a simple one-zone model that we will use to compare $\kappa${\tt monty} with the ray tracing codes {\tt RAPTOR} \footnote{https://github.com/jordydavelaar/raptor} and {\tt ipole-IL} and cross-compare the semi-analytical and numerical samplers in {\tt $\kappa$monty} and {\tt igrmonty}. The one-zone model is an isothermal sphere with a uniform density profile embedded in a uniform vertical magnetic field in flat spacetime. We solve the problem in spherical polar coordinates; the non-zero metric terms are given by
\begin{eqnarray}
g_{tt}&=&-1\\
g_{rr}&=& 1\\
g_{\theta\theta}&=& r^2\\
g_{\phi\phi}&=&r^2 \sin^2(\theta).
\end{eqnarray}
The plasma variables are given by,
\begin{eqnarray}
\rho &=& \rho_0 \\
\Theta_e &=& \Theta_{e,0}\\
B^t &=& 0.\\
B^r &=& B_0 \cos{\theta}\\
B^\theta &=& -B_0 \sin{\theta}/r\\
B^\phi &=& 0.
\end{eqnarray}
The solution is specified by constants $\rho_0$, $R_0$, $\Theta_{e,0}$, and $B_0$, along with an outer boundary to the domain $R_{\rm out}$. $\rho_0$ is given in terms of a characteristic Thomson depth $\tau_0$:
\begin{eqnarray}
\rho_0 = \frac{\tau_0}{\sigma_T R_0 \mathcal{L} \mathcal{N}}
\end{eqnarray}
where $\mathcal{L}$ is the code length unit conversion and $\mathcal{N}$ is the electron number density unit conversion. $B_0$ is expressed in terms of the plasma $\beta$ at $r=0$, $\beta_0$:
\begin{eqnarray}
B_0 = \sqrt{\frac{2 P_g}{\beta_0}}
\end{eqnarray}
where $P_g$ is the gas pressure. We set the adiabatic index to $\hat{\gamma}_{\rm adiab} = 13/9$, and the ratio between the proton and electron temperature is set to be $T_{\rm rat}=3.0$. 

\subsubsection{Comparison with RAPTOR}
To check whether the implementation of the emission coefficients is correct, we compute the SED without Compton scattering of the uniform sphere with $\kappa${\tt monty} and {\tt RAPTOR}. {\tt RAPTOR} \citep{bronzwaer2018,davelaar2018b,Bronzwaer2020} is a General Relativistic Ray tracing code that solves the covariant radiation transport equation in curved spacetime. GRRT methods are intrinsically different from MC methods since they use bundles of rays. This makes them ideal for computing synthetic images since only a small portion of the sky is covered by a camera. The camera consists of pixels, and every pixel is assigned an initial wavevector used to solve the geodesic equation backwards in time. Along these geodesics, the unpolarised radiation transport equation is solved.

\begin{figure}
  \centering
  \includegraphics[width=0.5\textwidth]{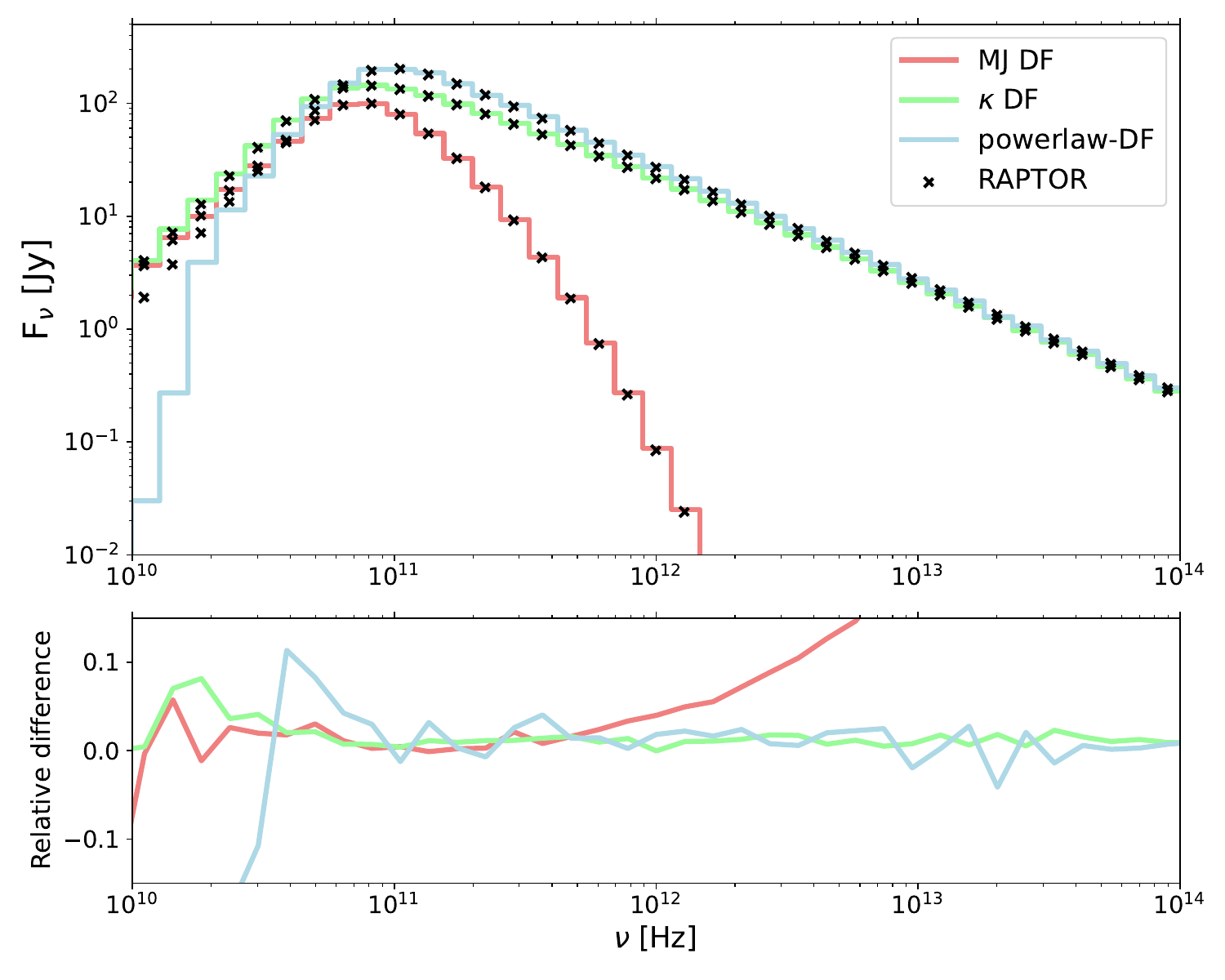}
  \caption{Comparison between $\kappa${\tt monty} and {\tt RAPTOR} for the uniform sphere test. Top panel: spectra for MJ-DF (red), $\kappa$-DF (green), and power-law DF (blue) for $\kappa$monty, black crosses data point from {
  \tt RAPTOR}. Bottom panel: the relative difference between the three DFs. The relative error is around 1\% for the majority of the SED. In the thermal case, at high frequency, the error grows. This is caused by a quickly shrinking emission region size, making sampling more difficult. All three models show slightly more noise at low frequencies due to the larger optical thickness. More superphotons are absorbed, which makes convergence more difficult compared to the optically thin part of the spectrum.}
  \label{fig:sphere-raptor}       
 \end{figure}

The setup and code-specific parameters for both $\kappa${\tt monty} as well as {\tt RAPTOR} are shown in Table~\ref{table:sphere}. The results of this test can be seen in Figure \ref{fig:sphere-raptor}. All three cases show minor discrepancies between the two codes except for the low-frequency part of the spectrum, which is dominated by MC noise. The thermal case shows deviations at the highest frequencies as well. The source becomes optically thin at these frequencies, and the emission region shrinks. The electrons responsible for this emission are at the exponential tail of the distribution functions, affecting the sampling. 

\begin{table}
\centering
\begin{tabular}{l l}
\hline
Source parameters  & value \\
\hline
Distance & $8.5 ~{\rm kpc}$\\
$M_{\rm BH}$ & $4.1\times10^6 M_\odot$ \\
\hline
Uniform sphere parameters  & \\
\hline
$R_0$ & 100\\
$\tau_0$ & $10^{-5}$\\
$\Theta_{e,0}$ &  10\\
$\beta_0$ & 20 \\
\hline
GRMHD parameters  & \\
\hline
$R_{\rm max}$ & 40\\
$\mathcal{M}$ & $10^{19}$ g\\
$T_{\rm rat}$ & 3.0 \\
\hline
DF parameters  & \\
\hline
$\kappa ~(p)$ & 4.0 (3.0) \\
$w$ & $\Theta_e \frac{\kappa-3}{\kappa}$ \\
$T_{\rm ratio}$ & 3.0 \\
$\nu_{\rm cutoff}$ & $5\times10^{13} ~{\rm Hz}$\\
$\gamma_{\rm cutoff}$ & $10^3$\\
$\gamma_{\rm min}$ & $25$\\
$\gamma_{\rm max}$ & $10^7$\\
\hline
Code parameters $\kappa${\tt monty}  & \\
\hline
$\nu_{\rm min}$ & $10^{9} ~{\rm Hz}$\\
$\nu_{\rm max}$ & $10^{16} ~{\rm Hz}$\\
$N_\theta$ & 180\\
$N_\phi$ & 90\\
${\rm RMAX}$ & $10^4$ $r_{\rm g}$\\
equatorial folding & no\\
\hline
Code parameters {\tt RAPTOR} and {\tt ipole-IL}
  & \\
\hline
field of view & 300 $r_{\rm g}$ \\
pixels & $512^2$\\
$r_{\rm cam}$ & $10^4$ $r_{\rm g}$\\
$i_{\rm cam}$ & $90^\circ$\\
\hline
\hline

\end{tabular}
\caption{Code and test parameters for all tests described in this paper.\label{table:sphere}}
\end{table}

 \begin{figure}
  \centering
  \includegraphics[width=0.5\textwidth]{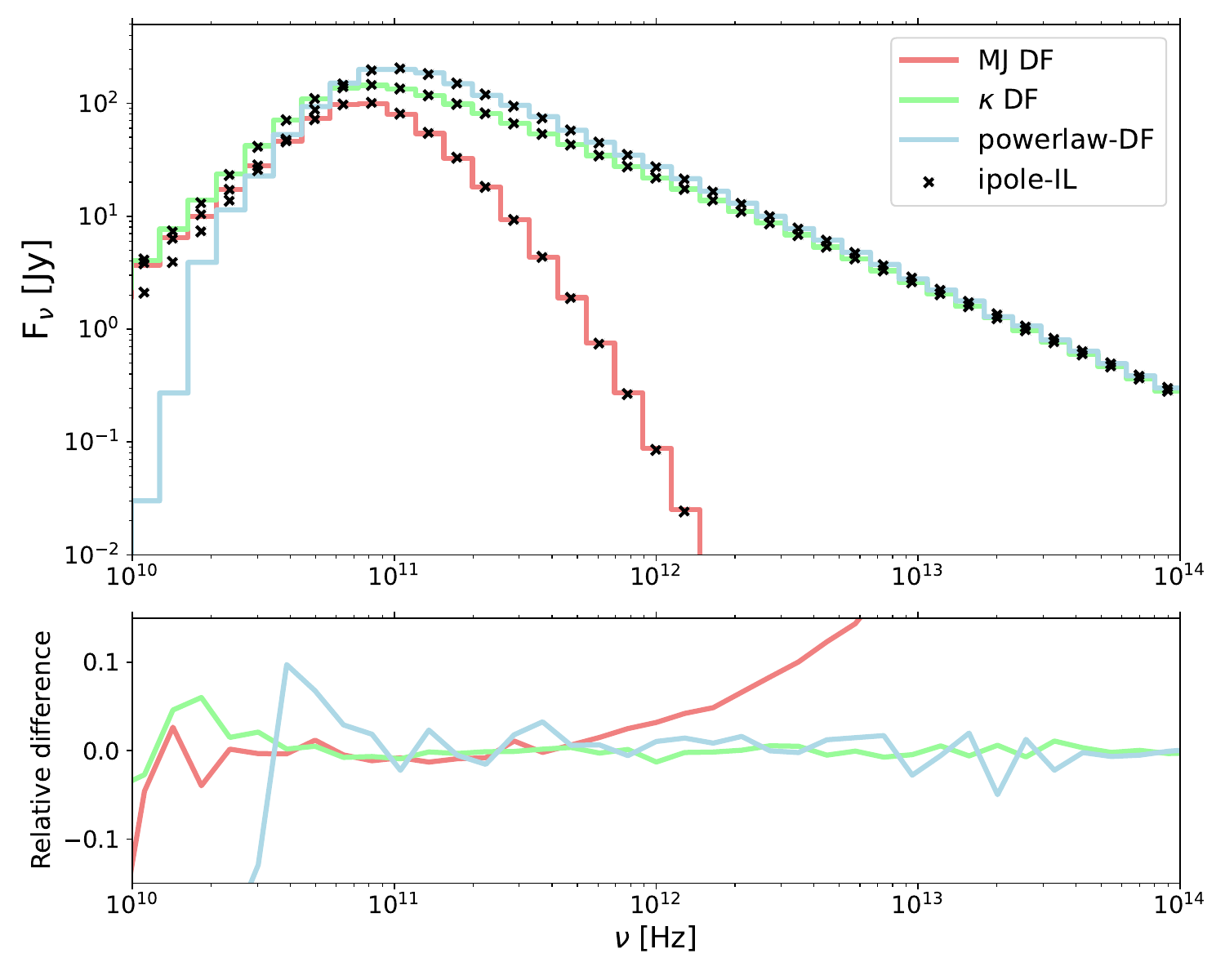}
  \caption{Comparison between $\kappa${\tt monty} and {\tt ipole-IL} for the uniform sphere test. Identical results compared to Figure \ref{fig:sphere-raptor}}
  \label{fig:sphere-ipole}       
 \end{figure}

 \subsubsection{Comparison with ipole-IL}

As an additional independent test to validate both $\kappa${\tt monty} as well as {\tt RAPTOR}, we also cross-compared the output of $\kappa${\tt monty} with the {\tt ipole-IL} ray tracing code \citep{moscibrodzka2018,wong2022}. This test is identical to the test with {\tt RAPTOR} presented in the previous subsection. The result can be seen in Figure \ref{fig:sphere-ipole}, and the agreement between the two codes is identical to the test with {\tt RAPTOR}.
 
\subsubsection{Compton Scattering test}

 \begin{figure*}
  \centering
  \includegraphics[width=\textwidth]{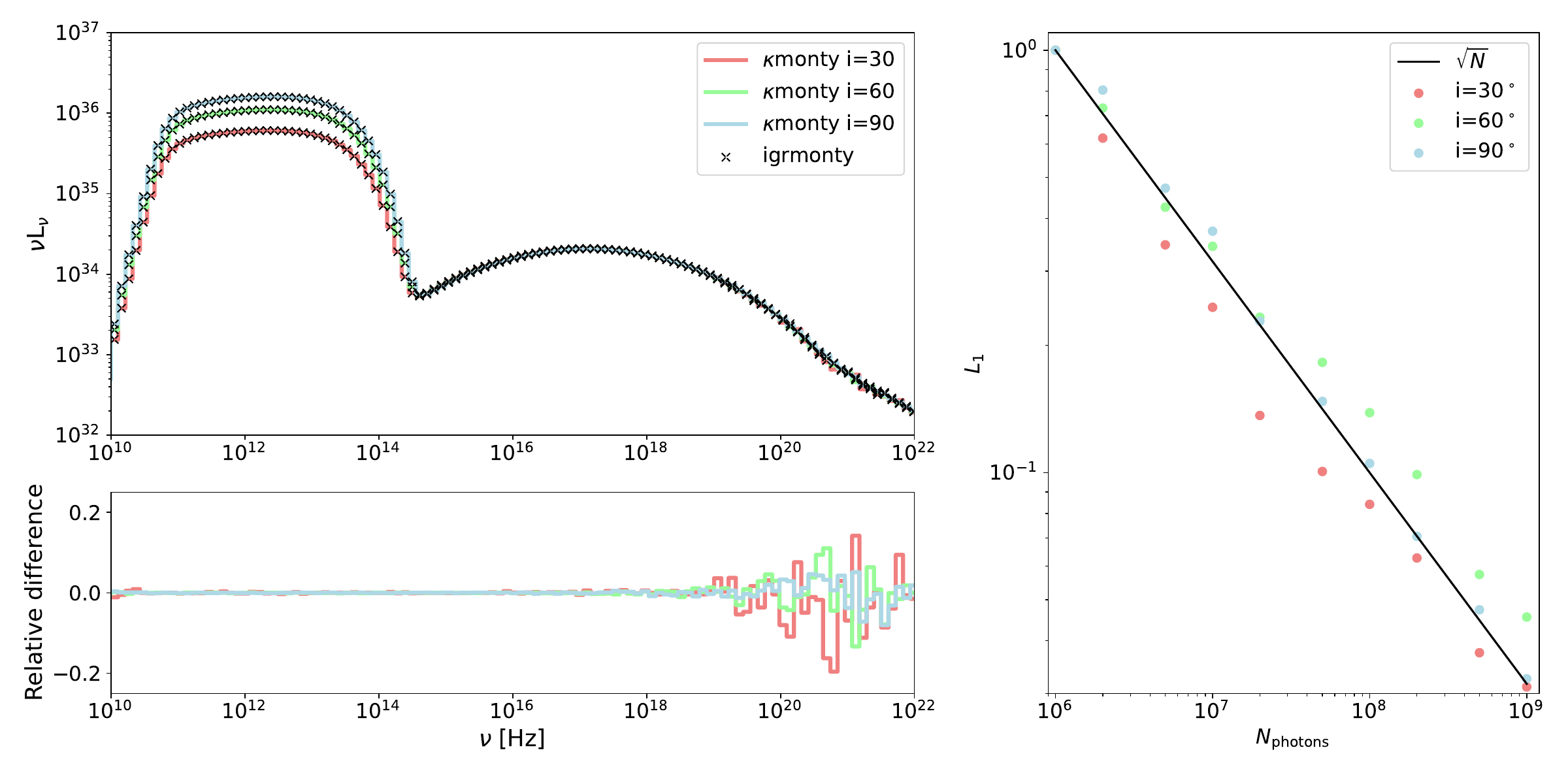}
  \caption{ Comparison between the semi-analytical ($\kappa$monty) and numerical sampler ({\tt igrmonty}) at three inclinations, top left: spectra at inclinations of $30^\circ$, $60^\circ$, and $90^\circ$. All three cases show consistent spectra with order 1\% differences (bottom left). The MC noise grows at high frequencies due to the low probability of double-scattering events. Right: $L_1$ convergence of the comparison between the semi-analytical and numerical samplers. As expected of MC methods, a clear $\sqrt{N}$ trend is visible. }
  \label{fig:sphere-spec}       
 \end{figure*}

To test the implementation of the non-thermal DFs within the Compton scattering module, we cross-compared a semi-analytical implementation with a numerical one in {\tt igrmonty}. The camera is positioned at a distance of $10^4 \mathcal{L}$, and we compute spectra for three inclinations $30^\circ, 60^\circ, 90^\circ$. The full set of parameters is similar to the ones shown in Table \ref{table:sphere}, except that $N_\theta=3$,$N_\phi=1$. And we use folding around the equator, meaning bins in the half-sphere above and below the equator are averaged to increase statistics.

The results of this comparison can be seen in Figure \ref{fig:sphere-spec}. There is clear consistency between the two methods, with a relative difference of less than 1\% for most of the SED. The MC noise grows at high frequencies due to the low probability of double-scattering events. To test the convergence of this test, we performed multiple runs with different amounts of initial superphotons. The convergence can be seen in the right panel of Figure \ref{fig:sphere-spec}. There is a clear $\sqrt{N}$ convergence visible, as would be expected of a Monte Carlo code.

 \subsection{GRMHD test}

A more challenging test is performed by comparing the synchrotron part of the SED between $\kappa${\tt monty} and {\tt RAPTOR} computed from a snapshot of a GRMHD simulation. The simulation, in CKS coordinates,
is the same as that presented in \citet{davelaar2019,Olivares2019}. The initial condition of this simulation is a \cite{fishbone1976} torus with black hole spin parameter $a_*=15/16$, inner radius $6 ~\rg$, pressure maximum at $12 ~\rg$, and adiabatic index $\hat{\gamma}=4/3$. The initial magnetic field profile is a single poloidal loop that follows isocontours of the density profile. The initial torus is weakly magnetised and set by the ratio between the maximum magnetic pressure $P_{\rm mag,max}$ and maximum gas pressure $P_{\rm max}$ and is set to be ${P_{\rm max}}/{P_{\rm mag,max}} = 100$.

Since the GRMHD simulation we used does not include electron thermodynamics, we use a parameterisation for the electron temperature

\begin{figure}
  \centering
  \includegraphics[width=0.5\textwidth]{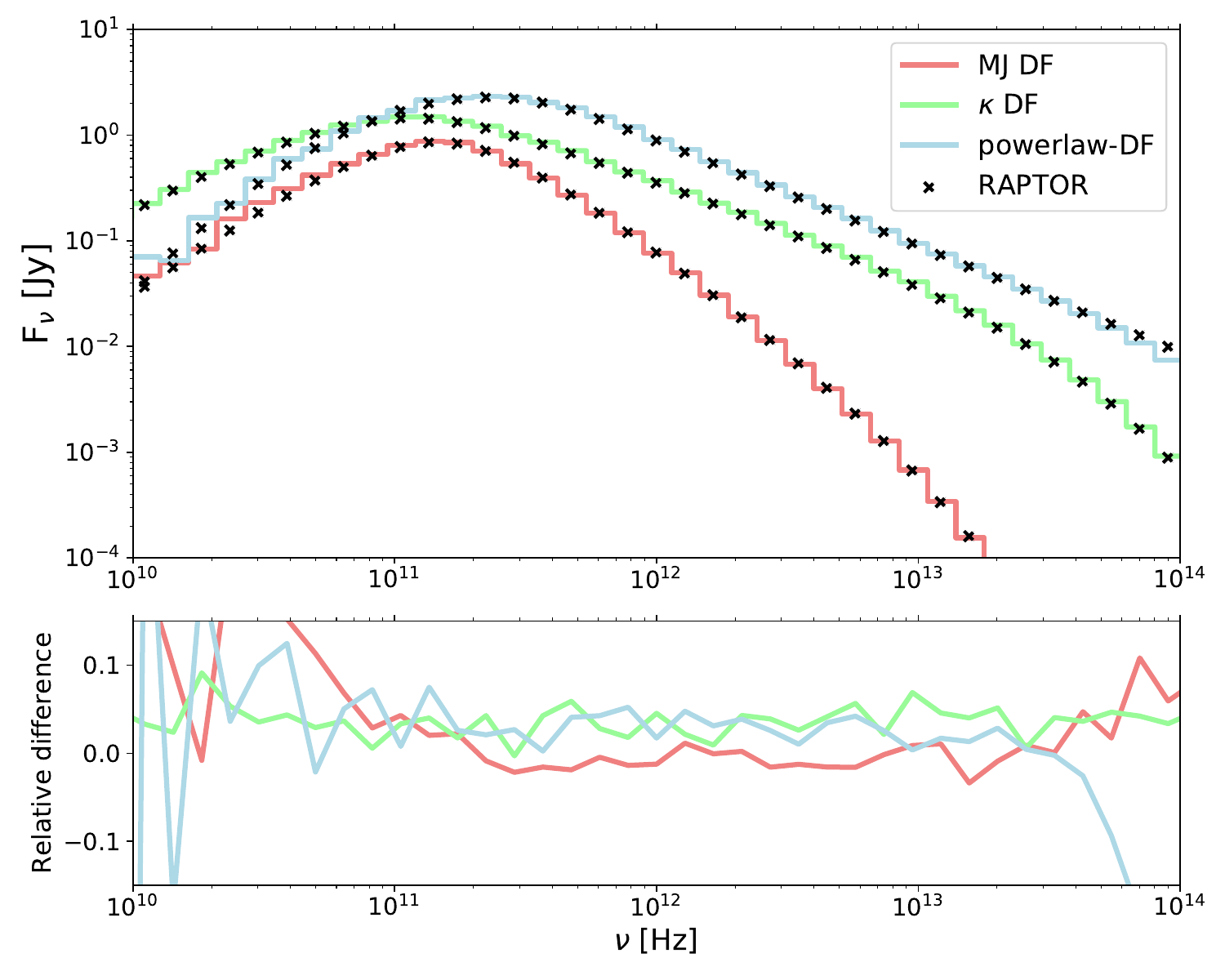}
  \caption{Comparison between {\tt RAPTOR} and $\kappa${\tt monty} for the GRMHD test, left: Thermal DF, middle: $\kappa$-DF, right: power-law DF. Consistent with the uniform sphere test, the deviations are of the order of 1\% per cent except for the high or low-frequency part of the spectrum.}
  \label{fig:raptor-test}       
 \end{figure}

 \begin{figure*}
  \centering
  \includegraphics[width=\textwidth]{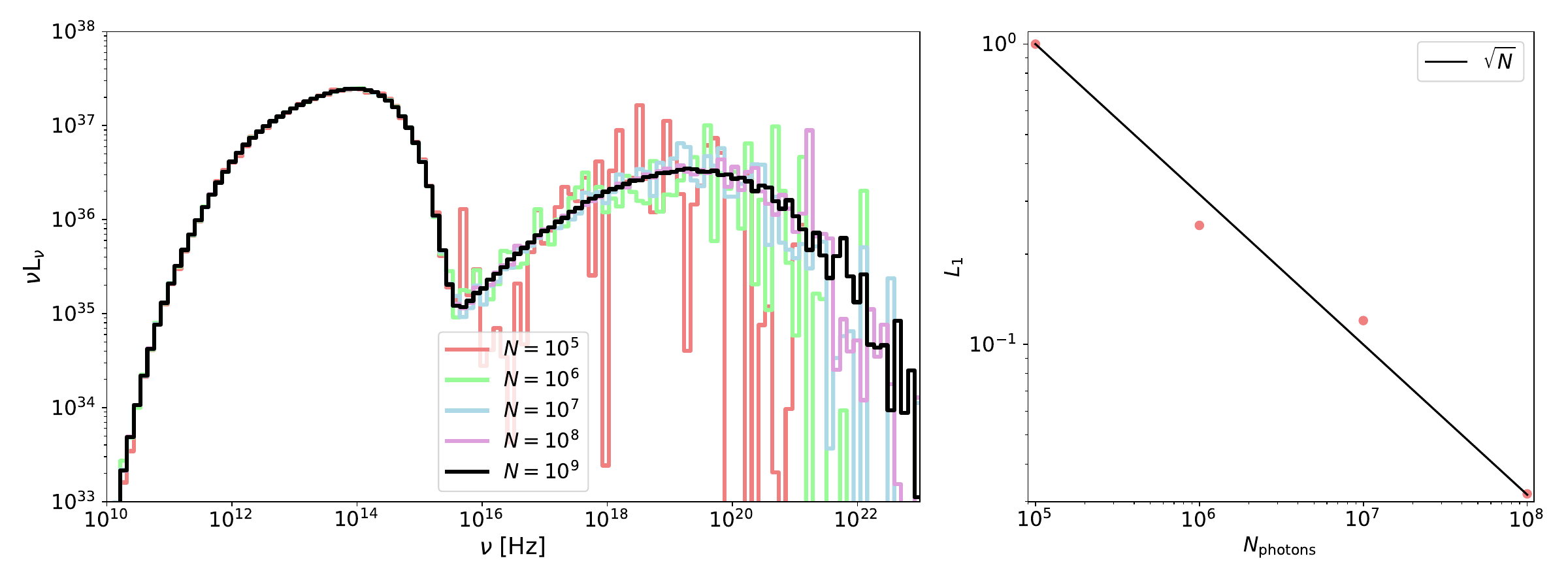}
  \caption{Self convergence of the spectra based on the GRMHD snapshot including Compton scattering, model uses the $\kappa$-DF. Left: spectra for $N=(10^5,10^6,10^7,10^8,10^9)$ superphotons. Right: self convergence of the solution with respect to the $N=10^9$ superphotons run. An evident $\sqrt{N}$ scaling is visible as expected for MC methods.}
  \label{fig:self-conv}       
 \end{figure*}

\begin{equation}
\Theta_{e}=\frac{U (\hat{\gamma}-1) m_{\rm p}}{\rho m_{\rm e}(T_{\rm rat}+1)},
\end{equation}
where $U$ is the internal energy, $m_{\rm p}$ the proton mass, $m_{\rm p}$ the electron mass, and $T_{\rm rat}$ the ratio between the proton to electron temperature which we set to $T_{\rm rat}=3.0$. To impose charge neutrality of the plasma, we set the electron number density equal to the proton number density.

The GRMHD simulation is scale-free to convert from code to c.g.s. units. Besides the aforementioned length and time unit, also a mass unit $\mathcal{M}$ is needed. The length and time units are given by the black hole length and gravitational timescales, $r_{\rm g} = G M_{\rm BH}/c^2$ and $t_{\rm g} = r_{\rm g}/c$, while the mass unit sets the energy content of the simulation and is tightly related to the mass accretion rate via $\dot{M}=\dot{M}_{\rm sim} \mathcal{M}/\mathcal{T}$. To convert the plasma variable to c.g.s. units the following conversion factors are used: $\rho_0 = \mathcal{M}/\mathcal{L}^3$,
$u_{0}=\rho_0 c^2$, and $B_0=c\sqrt{4\pi \rho_0}$.

The test-specific parameters for the camera, DF, and GRMHD parameters can be seen in Table \ref{table:sphere}. Only the inner $40 ~r_{\rm g}$ of the GRMHD domain is used to limit the field of view needed and speed up the convergence of the MC solution.

The results of this test can be seen in Figure \ref{fig:raptor-test}. For all three DFs, the error is close to 1\% in most frequency bins. For the thermal case at high frequency, the error grows at high frequencies, similar to the uniform sphere test. All three DFs show less agreement at lower frequencies (around $10^{10}$ Hz), due to opacity effects.

 \section{Code performance and availability}

Since the convergence of a Monte-Carlo simulation scales with $\sqrt{N}$, with $N$ the amount of superphotons, it is computationally demanding to acquire a fully converged solution. To accelerate the convergence, {\tt grmonty} was parallelized with {\tt OpenMP}, allowing it to run on multiple cores on a single node. To improve our code performance even further, we parallelized $\kappa${\tt monty} with {\tt MPI}, which allows us to run over many nodes. We identified two potential ways to {\tt MPI} parallelize our computations. One could either distribute the GRMHD domain over all the available {\tt MPI} processes and trace superphotons through the domain, this could lead to substantial communication overhead when superphotons leave/enter the domain of a processor, or would require a very labour intensive implementation where batched superphotons are send and received with non-blocking MPI. Alternatively, one could launch independent {\tt MPI} instances that all have the full domain in memory and at termination, sum all the resulting spectra. The first option has the benefit that it is memory efficient. However, scaling is limited to the IO overhead as the domain per {\tt MPI} instance gets smaller. The second option is more memory demanding but is trivial to implement and is easily scalable to large numbers of nodes as long as the domain fits within the memory per {\tt MPI} task. A typical $256^3$ simulations takes about five Gigabytes of memory, which allows for this implementation strategy, for high resolution simulations either the first method or a hydro {\tt Openmp+MPI} implementation should be explored. For $\kappa${\tt monty}, we opted for the second parallelization strategy. In the remainder of this section, we test our implementation for scalability and performance.

\begin{figure}
  \centering
  \includegraphics[width=0.5\textwidth]{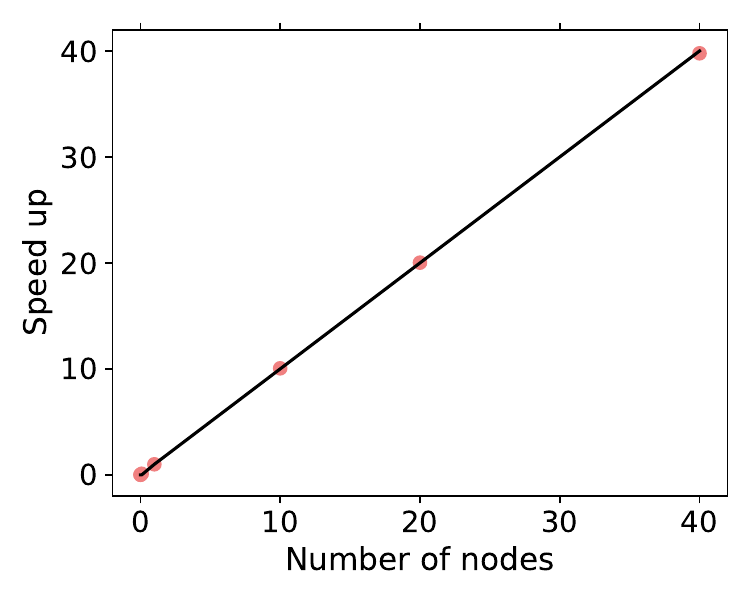}
  \caption{Scaling performance of the code for varying cores. We increased the amount of nodes from one to 40 nodes, each containing 128 cores, and computed the speed up. An evident linear scaling is visible in the red points, as the black line indicates perfect linear scaling.}
  \label{fig:scaling}       
 \end{figure}

The code is publicly available on GitHub\footnote{https://github.com/jordydavelaar/kmonty}. To test the performance of {\tt $\kappa$monty} we ran a {\tt BHAC} MKS GRMHD snapshot with scattering for the $\kappa$-DF and thermal-DF. First, we varied the number of initial superphotons to test for the solution's self-convergence. The resulting spectra for $N=(10^6,10^7,10^8,10^9)$ can be seen in the left panel of Figure \ref{fig:self-conv}. The right panel shows the convergence rate, which shows a $\sqrt{N}$ scaling. Secondly, we ran the code on varying amounts of nodes to test our code's scalability. For this test, we used nodes with 128 {\tt AMD Rome} cores with a total processing power per node of 4.6 teraflops. We compiled the code with the {\tt intel} compiler and standard {\tt intel} optimization flags. We varied the number of nodes from 1 to 40 nodes. Again, we use the GRMHD setup with scattering and start the code with $N=10^5$ superphotons per core. The scaling is shown in Figure \ref{fig:scaling}, and shows an evident linear scaling. Overall we achieve a speed of a few thousand superphotons per second per core, although note that this highly depends on the problem, stepsize, and scattering opacity. The 40 nodes run achieves a speed of ten million superphotons per second. The difference between the thermal, power-law, and $\kappa$-DFs is negligible since the computational bottleneck is the handling of the non-uniform data structure and geodesic integration. 

We also tested the performance of the semi-analytical samplers. We provide a stand-alone {\tt openmp} accelerated code of the sampling routines on GitHub \footnote{https://github.com/jordydavelaar/edf-samplers}. We ran the samplers on the same architecture as for the $\kappa${\tt monty} performance test. The $\kappa$-DF samples a few hundred thousand electrons per second per core, while the power-law and thermal DF samples around ten million electrons per second per core. Although the $\kappa$-DF is orders of magnitude slower, since it does not make use of heavily optimized {\tt gsl} samplers, the routine is only called at most one or two times per superphoton, meaning the computational cost is negligible when used by {\tt $\kappa$monty}.

 \section{Conclusion}\label{sec:concl}

We presented our new $\kappa${\tt monty} code. The code is an extension of {\tt grmonty} and now includes $\kappa$ and power-law distribution functions for both the radiative transfer coefficients and sampling routines. The code can also post-process the AMR data format of {\tt BHAC}. We tested our sampling routines by comparing the numerical output to the analytical DFs. We used a uniform isothermal sphere to test the implementations of the emission coefficients by comparing them with the ray-tracing code {\tt RAPTOR}. We tested the full emission and scattering kernels by comparing them to an implementation of the kappa distribution in {\tt igrmonty} that uses the numerical sampling routine. And finally, test the coupling to {\tt BHAC} by comparing the synchrotron emission with {\tt RAPTOR} by using a snapshot of a black hole simulation in Cartesian coordinates which uses AMR.

\section*{Acknowledgements}
The authors thank Oliver Porth, Sasha Philippov, Jeremy Schnittman, and Chi-kwan Chan for valuable discussions and feedback on the project. JD is supported by NASA grant NNX17AL82G and a Joint Columbia/Flatiron Postdoctoral Fellowship. Research at the Flatiron Institute is supported by the Simons Foundation. 
GNW is supported by the Taplin Fellowship. 
HO was supported by a Virtual Institute of
Accretion (VIA) postdoctoral fellowship from the Netherlands
Research School for Astronomy (NOVA).
This work has been assigned a document release number LA-UR-23-21157.
This work was partly funded by the ERC Synergy Grant ``BlackHoleCam-Imaging the Event Horizon of Black Holes'' (Grant 610058, \cite{goddi2017}). The GRMHD simulations were performed on the Dutch National Supercomputing cluster Cartesius and are funded by the NWO computing grant 16431. This research has made use of NASA's Astrophysics Data System.\\

{\it Software:} {\tt python} \citep{travis2007,jarrod2011}, {\tt scipy} \citep{jones2001}, {\tt numpy} \citep{walt2011}, {\tt matplotlib} \citep{hunter2007}, {\tt RAPTOR} \citep{bronzwaer2018,Bronzwaer2020}.

\section*{Data availability}
The data underlying this article will be shared on reasonable request to the corresponding author.

\label{lastpage}


\begin{thebibliography}{}

\bibitem[Baganoff et~al., 2003]{baganoff2003}
Baganoff, F.~K., Maeda, Y., Morris, M., Bautz, M., Brandt, W.~N., Cui, W.,
  Doty, J., Feigelson, E., Garmire, G., Pravdo, S., et~al. (2003).
\newblock Chandra x-ray spectroscopic imaging of sagittarius a* and the central
  parsec of the galaxy.
\newblock {\em \apj}, 591(2):891.

\bibitem[{Ball} et~al., 2016]{ball2016}
{Ball}, D., {{\"O}zel}, F., {Psaltis}, D., and {Chan}, C.-k. (2016).
\newblock {Particle Acceleration and the Origin of X-Ray Flares in GRMHD
  Simulations of SGR A}.
\newblock {\em \apj}, 826:77.

\bibitem[{B{\"o}ttcher} et~al., 2003]{bottcher2}
{B{\"o}ttcher}, M., {Jackson}, D.~R., and {Liang}, E.~P. (2003).
\newblock {Two-dimensional Monte Carlo/Fokker-Planck Simulations of Flares in
  Accretion Disk Corona Models}.
\newblock {\em apj}, 586:389--402.

\bibitem[{B{\"o}ttcher} and {Liang}, 2001]{bottcher1}
{B{\"o}ttcher}, M. and {Liang}, E.~P. (2001).
\newblock {Monte Carlo Simulations of Thermal-Nonthermal Radiation from a
  Neutron Star Magnetospheric Accretion Shell}.
\newblock {\em apj}, 552:248--258.

\bibitem[{Broderick} et~al., 2015]{broderick2015}
{Broderick}, A.~E., {Narayan}, R., {Kormendy}, J., {Perlman}, E.~S., {Rieke},
  M.~J., and {Doeleman}, S.~S. (2015).
\newblock {The Event Horizon of M87}.
\newblock {\em \apj}, 805:179.

\bibitem[{Bronzwaer} et~al., 2018]{bronzwaer2018}
{Bronzwaer}, T., {Davelaar}, J., {Younsi}, Z., {Mo{\'s}cibrodzka}, M.,
  {Falcke}, H., {Kramer}, M., and {Rezzolla}, L. (2018).
\newblock {RAPTOR. I. Time-dependent radiative transfer in arbitrary
  spacetimes}.
\newblock {\em \aap}, 613:A2.

\bibitem[{Bronzwaer} et~al., 2020]{Bronzwaer2020}
{Bronzwaer}, T., {Younsi}, Z., {Davelaar}, J., and {Falcke}, H. (2020).
\newblock {RAPTOR II: Polarized radiative transfer in curved spacetime}.
\newblock {\em arXiv e-prints}, page arXiv:2007.03045.

\bibitem[{Canfield} et~al., 1987]{canfield}
{Canfield}, E., {Howard}, W.~M., and {Liang}, E.~P. (1987).
\newblock {Inverse Comptonization by one-dimensional relativistic electrons}.
\newblock {\em \apj}, 323:565--574.

\bibitem[{Chael} et~al., 2017]{Chael2017}
{Chael}, A.~A., {Narayan}, R., and {Sadowski}, A. (2017).
\newblock {Evolving non-thermal electrons in simulations of black hole
  accretion}.
\newblock {\em \mnras}, 470(2):2367--2386.

\bibitem[{Chan} et~al., 2009]{Chan2009}
{Chan}, C.-k., {Liu}, S., {Fryer}, C.~L., {Psaltis}, D., {{\"O}zel}, F.,
  {Rockefeller}, G., and {Melia}, F. (2009).
\newblock {MHD Simulations of Accretion onto Sgr A*: Quiescent Fluctuations,
  Outbursts, and Quasiperiodicity}.
\newblock {\em \apj}, 701(1):521--534.

\bibitem[{Chan} et~al., 2015a]{Chan2015b}
{Chan}, C.-k., {Psaltis}, D., {{\"O}zel}, F., {Medeiros}, L., {Marrone}, D.,
  {Sadowski}, A., and {Narayan}, R. (2015a).
\newblock {Fast Variability and Millimeter/IR Flares in GRMHD Models of Sgr A*
  from Strong-field Gravitational Lensing}.
\newblock {\em \apj}, 812(2):103.

\bibitem[{Chan} et~al., 2015b]{Chan2015a}
{Chan}, C.-K., {Psaltis}, D., {{\"O}zel}, F., {Narayan}, R., and {Sadowski}, A.
  (2015b).
\newblock {The Power of Imaging: Constraining the Plasma Properties of GRMHD
  Simulations using EHT Observations of Sgr A*}.
\newblock {\em \apj}, 799(1):1.

\bibitem[{Chatterjee} et~al., 2020]{chatterjee2020}
{Chatterjee}, K., {Markoff}, S., {Neilsen}, J., {Younsi}, Z., {Witzel}, G.,
  {Tchekhovskoy}, A., {Yoon}, D., {Ingram}, A., {van der Klis}, M., {Boyce},
  H., {Do}, T., {Haggard}, D., and {Nowak}, M. (2020).
\newblock {General relativistic MHD simulations of non-thermal flaring in
  Sagittarius A*}.
\newblock {\em arXiv e-prints}, page arXiv:2011.08904.

\bibitem[{Cruz-Osorio} et~al., 2022]{ACO2022}
{Cruz-Osorio}, A., {Fromm}, C.~M., {Mizuno}, Y., {Nathanail}, A., {Younsi}, Z.,
  {Porth}, O., {Davelaar}, J., {Falcke}, H., {Kramer}, M., and {Rezzolla}, L.
  (2022).
\newblock {State-of-the-art energetic and morphological modelling of the
  launching site of the M87 jet}.
\newblock {\em Nature Astronomy}, 6:103--108.

\bibitem[{Davelaar} et~al., 2018a]{davelaar2018b}
{Davelaar}, J., {Bronzwaer}, T., {Kok}, D., {Younsi}, Z., {Mo{\'s}cibrodzka},
  M., and {Falcke}, H. (2018a).
\newblock {Observing supermassive black holes in virtual reality}.
\newblock {\em arXiv e-prints}, page arXiv:1811.08369.

\bibitem[{Davelaar} et~al., 2018b]{davelaar2018}
{Davelaar}, J., {Mo{\'s}cibrodzka}, M., {Bronzwaer}, T., and {Falcke}, H.
  (2018b).
\newblock {General relativistic magnetohydrodynamical {\ensuremath{\kappa}}-jet
  models for Sagittarius A*}.
\newblock {\em \aap}, 612:A34.

\bibitem[{Davelaar} et~al., 2019]{davelaar2019}
{Davelaar}, J., {Olivares}, H., {Porth}, O., {Bronzwaer}, T., {Janssen}, M.,
  {Roelofs}, F., {Mizuno}, Y., {Fromm}, C.~M., {Falcke}, H., and {Rezzolla}, L.
  (2019).
\newblock {Modeling non-thermal emission from the jet-launching region of M 87
  with adaptive mesh refinement}.
\newblock {\em arXiv e-prints}, page arXiv:1906.10065.

\bibitem[{Decker} and {Krimigis}, 2003]{decker}
{Decker}, R.~B. and {Krimigis}, S.~M. (2003).
\newblock {Voyager observations of low-energy ions during solar cycle 23}.
\newblock {\em Advances in Space Research}, 32:597--602.

\bibitem[{Dolence} et~al., 2009]{grmonty}
{Dolence}, J.~C., {Gammie}, C.~F., {Mo{\'s}cibrodzka}, M., and {Leung}, P.~K.
  (2009).
\newblock {grmonty: A Monte Carlo Code for Relativistic Radiative Transport}.
\newblock {\em \apjs}, 184:387--397.

\bibitem[{Dolence} et~al., 2012]{dolence}
{Dolence}, J.~C., {Gammie}, C.~F., {Shiokawa}, H., and {Noble}, S.~C. (2012).
\newblock {Near-infrared and X-Ray Quasi-periodic Oscillations in Numerical
  Models of Sgr A*}.
\newblock {\em apjl}, 746:L10.

\bibitem[Eckart et~al., 2004]{eckart2004}
Eckart, A., Baganoff, F., Morris, M., Bautz, M., Brandt, W.~N., Garmire, G.,
  Genzel, R., Ott, T., Ricker, G., Straubmeier, C., et~al. (2004).
\newblock First simultaneous nir/x-ray detection of a flare from sgr a.
\newblock {\em \aap}, 427(1):1--11.

\bibitem[{EHT Collaboration} et~al., 2019a]{eht-paperI}
{EHT Collaboration} et~al. (2019a).
\newblock {\em ApJL}.
\newblock 875, L1 (Paper I).

\bibitem[{EHT Collaboration} et~al., 2019b]{eht-paperV}
{EHT Collaboration} et~al. (2019b).
\newblock {\em ApJL}.
\newblock 875, L5 (Paper V).

\bibitem[{Event Horizon Telescope Collaboration}, 2022]{ehtsgrapaperV}
{Event Horizon Telescope Collaboration} (2022).
\newblock {First Sagittarius A* Event Horizon Telescope Results. V. Testing
  Astrophysical Models of the Galactic Center Black Hole}.
\newblock {\em \apjl}, 930(2):L16.

\bibitem[Fishbone and Moncrief, 1976]{fishbone1976}
Fishbone, L.~G. and Moncrief, V. (1976).
\newblock Relativistic fluid disks in orbit around {{Kerr}} black holes.
\newblock {\em The Astrophysical Journal}, 207:962--976.

\bibitem[{Fromm} et~al., 2022]{Fromm2022}
{Fromm}, C.~M., {Cruz-Osorio}, A., {Mizuno}, Y., {Nathanail}, A., {Younsi}, Z.,
  {Porth}, O., {Olivares}, H., {Davelaar}, J., {Falcke}, H., {Kramer}, M., and
  {Rezzolla}, L. (2022).
\newblock {Impact of non-thermal particles on the spectral and structural
  properties of M87}.
\newblock {\em \aap}, 660:A107.

\bibitem[{Goddi} et~al., 2017]{goddi2017}
{Goddi}, C., {Falcke}, H., {Kramer}, M., {Rezzolla}, L., {Brinkerink}, C.,
  {Bronzwaer}, T., {Davelaar}, J.~R.~J., {Deane}, R., {de Laurentis}, M.,
  {Desvignes}, G., {Eatough}, R.~P., {Eisenhauer}, F., {Fraga-Encinas}, R.,
  {Fromm}, C.~M., {Gillessen}, S., {Grenzebach}, A., {Issaoun}, S.,
  {Jan{\ss}en}, M., {Konoplya}, R., {Krichbaum}, T.~P., {Laing}, R., {Liu}, K.,
  {Lu}, R.~S., {Mizuno}, Y., {Moscibrodzka}, M., {M{\"u}ller}, C., {Olivares},
  H., {Pfuhl}, O., {Porth}, O., {Roelofs}, F., {Ros}, E., {Schuster}, K.,
  {Tilanus}, R., {Torne}, P., {van Bemmel}, I., {van Langevelde}, H.~J., {Wex},
  N., {Younsi}, Z., and {Zhidenko}, A. (2017).
\newblock {BlackHoleCam: Fundamental physics of the galactic center}.
\newblock {\em International Journal of Modern Physics D}, 26:1730001--239.

\bibitem[{Harris} et~al., 2009]{Harris2009}
{Harris}, D.~E., {Cheung}, C.~C., {Stawarz}, {\L}., {Biretta}, J.~A., and
  {Perlman}, E.~S. (2009).
\newblock {Variability Timescales in the M87 Jet: Signatures of E $^{2}$
  Losses, Discovery of a Quasi Period in HST-1, and the Site of TeV Flaring}.
\newblock {\em \apj}, 699(1):305--314.

\bibitem[{Hunter}, 2007]{hunter2007}
{Hunter}, J.~D. (2007).
\newblock {Matplotlib: A 2D Graphics Environment}.
\newblock {\em Computing in Science and Engineering}, 9:90--95.

\bibitem[Jones et~al., 2001]{jones2001}
Jones, E., Oliphant, T., Peterson, P., et~al. (2001).
\newblock {SciPy}: Open source scientific tools for {Python}.
\newblock [Online].

\bibitem[{Kerr}, 1963]{kerr1963}
{Kerr}, R.~P. (1963).
\newblock {Gravitational Field of a Spinning Mass as an Example of
  Algebraically Special Metrics}.
\newblock {\em \prl}, 11:237--238.

\bibitem[{Kunz} et~al., 2016]{kunz2016}
{Kunz}, M.~W., {Stone}, J.~M., and {Quataert}, E. (2016).
\newblock {Magnetorotational Turbulence and Dynamo in a Collisionless Plasma}.
\newblock {\em Physical Review Letters}, 117(23):235101.

\bibitem[{Laurent} and {Titarchuk}, 1999]{laurent}
{Laurent}, P. and {Titarchuk}, L. (1999).
\newblock {The Converging Inflow Spectrum Is an Intrinsic Signature for a Black
  Hole: Monte Carlo Simulations of Comptonization on Free-falling Electrons}.
\newblock {\em apj}, 511:289--297.

\bibitem[{Leung} et~al., 2011a]{leung}
{Leung}, P.~K., {Gammie}, C.~F., and {Noble}, S.~C. (2011a).
\newblock {Numerical Calculation of Magnetobremsstrahlung Emission and
  Absorption Coefficients}.
\newblock {\em \apj}, 737:21.

\bibitem[{Leung} et~al., 2011b]{leung2011}
{Leung}, P.~K., {Gammie}, C.~F., and {Noble}, S.~C. (2011b).
\newblock {Numerical Calculation of Magnetobremsstrahlung Emission and
  Absorption Coefficients}.
\newblock {\em \apj}, 737:21.

\bibitem[Livadiotis and McComas, 2013]{Livadiotis2013}
Livadiotis, G. and McComas, D.~J. (2013).
\newblock Understanding kappa distributions: A toolbox for space science and
  astrophysics.
\newblock {\em Space Science Reviews}, 175(1):183--214.

\bibitem[{Mao} et~al., 2017]{mao2016}
{Mao}, S.~A., {Dexter}, J., and {Quataert}, E. (2017).
\newblock {The impact of non-thermal electrons on event horizon scale images
  and spectra of Sgr A*}.
\newblock {\em \mnras}, 466(4):4307--4319.

\bibitem[Marshall et~al., 2002]{Marshall2002}
Marshall, H.~L., Miller, B.~P., Davis, D.~S., Perlman, E.~S., Wise, M.,
  Canizares, C.~R., and Harris, D.~E. (2002).
\newblock A high-resolution x-ray image of the jet in m87.
\newblock {\em \apj}, 564(2):683--687.

\bibitem[Millman and Aivazis, 2011]{jarrod2011}
Millman, K.~J. and Aivazis, M. (2011).
\newblock Python for scientists and engineers.
\newblock {\em Computing in Science \& Engineering}, 13(2):9--12.

\bibitem[{Mo{\'s}cibrodzka}, 2020]{Moscibrodzka2020}
{Mo{\'s}cibrodzka}, M. (2020).
\newblock {General relativistic polarized radiative transfer with
  inverse-Compton scatterings}.
\newblock {\em \mnras}, 491(4):4807--4815.

\bibitem[{Mo{\'s}cibrodzka} et~al., 2016]{moscibrodzka2016}
{Mo{\'s}cibrodzka}, M., {Falcke}, H., and {Shiokawa}, H. (2016).
\newblock {General relativistic magnetohydrodynamical simulations of the jet in
  M 87}.
\newblock {\em \aap}, 586:A38.

\bibitem[{Mo{\'s}cibrodzka} and {Gammie}, 2018]{moscibrodzka2018}
{Mo{\'s}cibrodzka}, M. and {Gammie}, C.~F. (2018).
\newblock {IPOLE - semi-analytic scheme for relativistic polarized radiative
  transport}.
\newblock {\em \mnras}, 475(1):43--54.

\bibitem[{Mo{\'s}cibrodzka}, 2022]{mosc2022}
{Mo{\'s}cibrodzka}, M.~A. (2022).
\newblock {Polarization-sensitive Compton Scattering by Accelerated Electrons}.
\newblock {\em \apjs}, 263(1):6.

\bibitem[{Narayan} et~al., 2016]{narayan}
{Narayan}, R., {Zhu}, Y., {Psaltis}, D., and {Sa{\c d}owski}, A. (2016).
\newblock {HEROIC: 3D general relativistic radiative post-processor with
  comptonization for black hole accretion discs}.
\newblock {\em mnras}, 457:608--628.

\bibitem[{Nathanail} et~al., 2020]{Nathanail2020}
{Nathanail}, A., {Fromm}, C.~M., {Porth}, O., {Olivares}, H., {Younsi}, Z.,
  {Mizuno}, Y., and {Rezzolla}, L. (2020).
\newblock {Plasmoid formation in global GRMHD simulations and AGN flares}.
\newblock {\em \mnras}, 495(2):1549--1565.

\bibitem[Oliphant, 2007]{travis2007}
Oliphant, T.~E. (2007).
\newblock Python for scientific computing.
\newblock {\em Computing in Science \& Engineering}, 9(3):10--20.

\bibitem[{Olivares} et~al., 2019]{Olivares2019}
{Olivares}, H., {Porth}, O., {Davelaar}, J., {Most}, E.~R., {Fromm}, C.~M.,
  {Mizuno}, Y., {Younsi}, Z., and {Rezzolla}, L. (2019).
\newblock {Constrained transport and adaptive mesh refinement in the Black Hole
  Accretion Code}.
\newblock {\em arXiv e-prints}, page arXiv:1906.10795.

\bibitem[{{\"O}zel} et~al., 2000]{ozel}
{{\"O}zel}, F., {Psaltis}, D., and {Narayan}, R. (2000).
\newblock {Hybrid Thermal-Nonthermal Synchrotron Emission from Hot Accretion
  Flows}.
\newblock {\em \apj}, 541:234--249.

\bibitem[{Pandya} et~al., 2016]{pandya2016}
{Pandya}, A., {Zhang}, Z., {Chandra}, M., and {Gammie}, C.~F. (2016).
\newblock {Polarized Synchrotron Emissivities and Absorptivities for
  Relativistic Thermal, Power-law, and Kappa Distribution Functions}.
\newblock {\em \apj}, 822:34.

\bibitem[Perlman and Wilson, 2005]{perlman2005}
Perlman, E.~S. and Wilson, A.~S. (2005).
\newblock The x-ray emissions from the m87 jet: Diagnostics and physical
  interpretation.
\newblock {\em \apj}, 627(1):140.

\bibitem[{Porth} et~al., 2017]{porth2017}
{Porth}, O., {Olivares}, H., {Mizuno}, Y., {Younsi}, Z., {Rezzolla}, L.,
  {Moscibrodzka}, M., {Falcke}, H., and {Kramer}, M. (2017).
\newblock {The black hole accretion code}.
\newblock {\em Computational Astrophysics and Cosmology}, 4:1.

\bibitem[{Prieto} et~al., 2016]{prieto2016}
{Prieto}, M.~A., {Fern{\'a}ndez-Ontiveros}, J.~A., {Markoff}, S., {Espada}, D.,
  and {Gonz{\'a}lez-Mart{\'\i}n}, O. (2016).
\newblock {The central parsecs of M87: jet emission and an elusive accretion
  disc}.
\newblock {\em \mnras}, 457:3801--3816.

\bibitem[{Quataert}, 2004]{Quataert2004}
{Quataert}, E. (2004).
\newblock {A Dynamical Model for Hot Gas in the Galactic Center}.
\newblock {\em \apj}, 613(1):322--325.

\bibitem[{Ripperda} et~al., 2020]{Ripperda2020}
{Ripperda}, B., {Bacchini}, F., and {Philippov}, A.~A. (2020).
\newblock {Magnetic Reconnection and Hot Spot Formation in Black Hole Accretion
  Disks}.
\newblock {\em \apj}, 900(2):100.

\bibitem[{Ripperda} et~al., 2022]{Ripperda2022}
{Ripperda}, B., {Liska}, M., {Chatterjee}, K., {Musoke}, G., {Philippov},
  A.~A., {Markoff}, S.~B., {Tchekhovskoy}, A., and {Younsi}, Z. (2022).
\newblock {Black Hole Flares: Ejection of Accreted Magnetic Flux through 3D
  Plasmoid-mediated Reconnection}.
\newblock {\em \apjl}, 924(2):L32.

\bibitem[{Ryan} et~al., 2015]{ryan}
{Ryan}, B.~R., {Dolence}, J.~C., and {Gammie}, C.~F. (2015).
\newblock {bhlight: General Relativistic Radiation Magnetohydrodynamics with
  Monte Carlo Transport}.
\newblock {\em apj}, 807:31.

\bibitem[{Schnittman} and {Krolik}, 2009]{schnittman2}
{Schnittman}, J.~D. and {Krolik}, J.~H. (2009).
\newblock {X-ray Polarization from Accreting Black Holes: The Thermal State}.
\newblock {\em apj}, 701:1175--1187.

\bibitem[{Schnittman} et~al., 2006]{schnittman1}
{Schnittman}, J.~D., {Krolik}, J.~H., and {Hawley}, J.~F. (2006).
\newblock {Light Curves from an MHD Simulation of a Black Hole Accretion Disk}.
\newblock {\em apj}, 651:1031--1048.

\bibitem[{Stern} et~al., 1995]{stern}
{Stern}, B.~E., {Begelman}, M.~C., {Sikora}, M., and {Svensson}, R. (1995).
\newblock {A large-particle Monte Carlo code for simulating non-linear
  high-energy processes near compact objects}.
\newblock {\em mnras}, 272:291--307.

\bibitem[{van der Walt} et~al., 2011]{walt2011}
{van der Walt}, S., {Colbert}, S.~C., and {Varoquaux}, G. (2011).
\newblock {The NumPy Array: A Structure for Efficient Numerical Computation}.
\newblock {\em Computing in Science and Engineering}, 13(2):22--30.

\bibitem[Wilson and Yang, 2001]{Wilson2001}
Wilson, S. and Yang, Y. (2001).
\newblock Chandra x‐ray imaging and spectroscopy of the m87 jet and nucleus.
\newblock {\em \apj}, 568.

\bibitem[{Wong} et~al., 2022]{wong2022}
{Wong}, G.~N., {Prather}, B.~S., {Dhruv}, V., {Ryan}, B.~R.,
  {Mo{\'s}cibrodzka}, M., {Chan}, C.-k., {Joshi}, A.~V., {Yarza}, R.,
  {Ricarte}, A., {Shiokawa}, H., {Dolence}, J.~C., {Noble}, S.~C., {McKinney},
  J.~C., and {Gammie}, C.~F. (2022).
\newblock {PATOKA: Simulating Electromagnetic Observables of Black Hole
  Accretion}.
\newblock {\em \apjs}, 259(2):64.

\bibitem[Xiao, 2006]{xiao2006}
Xiao, F. (2006).
\newblock Modelling energetic particles by a relativistic kappa-loss-cone
  distribution function in plasmas.
\newblock {\em Plasma Physics and Controlled Fusion}, 48(2):203.

\bibitem[{Yao} et~al., 2005]{yao}
{Yao}, Y., {Zhang}, S.~N., {Zhang}, X., {Feng}, Y., and {Robinson}, C.~R.
  (2005).
\newblock {Studying the Properties of Accretion Disks and Coronae in Black Hole
  X-Ray Binaries with Monte Carlo Simulation}.
\newblock {\em apj}, 619:446--454.

\bibitem[{Yuan} et~al., 2003]{yuan2003}
{Yuan}, F., {Quataert}, E., and {Narayan}, R. (2003).
\newblock {Nonthermal Electrons in Radiatively Inefficient Accretion Flow
  Models of Sagittarius A*}.
\newblock {\em \apj}, 598:301--312.

\bibitem[{Zhang} et~al., 2019]{zhang2019}
{Zhang}, W., {Dov{\v{c}}iak}, M., and {Bursa}, M. (2019).
\newblock {Constraining the Size of the Corona with Fully Relativistic
  Calculations of Spectra of Extended Coronae. I. The Monte Carlo Radiative
  Transfer Code}.
\newblock {\em \apj}, 875(2):148.

\end{thebibliography}
\end{document}